\documentclass[useAMS,usenatbib]{mn2e}

\usepackage{graphicx}
\usepackage{lscape}

\title[$\delta$ Sct components in binary systems]{Survey for $\delta$ Sct components in eclipsing binaries and new correlations between pulsation frequency and fundamental stellar characteristics}
\author[A. Liakos, P. Niarchos, E. Soydugan and P. Zasche]{A. Liakos$^{1}$\thanks{E-mail:
alliakos@phys.uoa.gr}, P. Niarchos$^{1}$, E. Soydugan$^{2}$ and P. Zasche$^{3}$\\
$^{1}$Dept. of Astrophysics, Astronomy and Mechanics, University of Athens, GR-157 84, Zografos, Athens, Hellas\\
$^{2}$Dept. of Physics, Faculty of Arts and Sciences, \c{C}anakkale Onsekiz Mart University \& \c{C}anakkale Onsekiz Mart University Observatory,\\
Terzio\v{g}lu Campus, TR-17020, \c{C}anakkale, Turkey\\
$^{3}$Astronomical Institute, Faculty of Mathematics and Physics, Charles University Prague, CZ-180 00 Praha 8,\\
V Hole\v{s}ovi\v{c}k\'ach 2, Czech Republic}

\begin{document}

\date{Accepted 201X December XX. Received 201X December XX; in original form 201X October XX}

\pagerange{\pageref{firstpage}--\pageref{lastpage}} \pubyear{201X}

\maketitle

\label{firstpage}

\begin{abstract}
CCD observations of 68 eclipsing binary systems, candidates for containing $\delta$~Scuti components, were obtained. Their light curves are analyzed using the Period04 software for possible pulsational behaviour. For the systems QY~Aql, CZ~Aqr, TY~Cap, WY~Cet, UW~Cyg, HL~Dra, HZ~Dra, AU~Lac, CL~Lyn and IO~UMa complete light curves were observed due to the detection of a pulsating component. All of them, except QY~Aql and IO~UMa, are analysed with modern astronomical softwares in order to determine their geometrical and pulsational characteristics. Spectroscopic observations of WY~Cet and UW~Cyg were used to estimate the spectral class of their primary components, while for HZ~Dra radial velocities of its primary were measured. O$-$C diagram analysis was performed for the cases showing peculiar orbital period variations, namely CZ~Aqr, TY~Cap, WY~Cet and UW~Cyg, with the aim of obtaining a comprehensive picture of these systems. An updated catalogue of 74 close binaries including a $\delta$~Scuti companion is presented. Moreover, a connection between orbital and pulsation periods, as well as a correlation between evolutionary status and dominant pulsation frequency for these systems is discussed.
\end{abstract}


\section{Introduction}

In general, close binaries, and especially the eclipsing ones, are stellar objects whose absolute parameters and evolutionary status can be easily derived from observations. Single $\delta$~Scuti stars show discrepancy from binary-members regarding their evolutionary status. The single ones are situated on the Main Sequence (hereafter MS) or moving from it to the giant branch. On the other hand, the $\delta$~Scuti components in binaries show slow evolution through the MS, thus they are very useful tools to diagnose this extraordinary part of a stellar lifetime. This difference in evolution is connected with the mass transfer process and tidal distortions occurring in binary systems during their MS life \citep{MK03,SO06a}. Therefore, single and binary-contained $\delta$~Scuti stars, although they present similar pulsational properties, should not be considered of the same type due to a likely different evolutionary past. Especially the ones in classical Algols show variable pulsational characteristics due to mass gain \citep{MK04,MK07}, a process that is responsible for differences in the excitation mechanism, i.e. $\kappa$-driven oscillations compared to tidally induced or mass-accretion induced oscillations. Additionally, calculation of the absolute parameters and the identification of the oscillating characteristics of a binary's pulsational component provide the means to obtain a detailed picture of the star. Obviously, the larger the sample of such stars the more information and conclusions can be derived.

During the last decade interest for pulsating stars in close binaries has increased significantly and a lot of discoveries announced. \citet{MK04} introduced the oEA (oscillating EA) stars as the (B)A-F spectral type mass-accreting MS pulsators in semi-detached Algol-type eclipsing binary (hereafter EBs) systems. \citet{SO06a} made a first attempt to find a connection between pulsation and orbital periods of systems with $\delta$~Scuti component and resulted into a linear relation. \citet{ZH10} published a catalogue containing 89 systems and distinguished them according to their pulsational properties. \citet{SO11} also published a similar list including 43 cases of systems including a $\delta$~Sct component. Space missions such as \textit{CoRot} and \textit{Kepler} have been discovering many pulsating stars in binaries (cf. \citealt{SOU11}; \citealt{DA10}). Their measurements provide the means to derive many pulsation frequencies and identify the oscillating modes with unprecedented accuracy.

The present work is a continuation of the survey for $\delta$~Scuti components in EBs of \citet{LN09}. Eight new such systems are presented herein for the first time, while the results of some cases that were recently discovered by other investigators are confirmed. The majority of the observed systems were selected from the lists of \citet{SO06b}. The results of both surveys, which enriched significantly ($\sim14\%$) the sample of such binaries, can therefore be used for future space and/or ground based observing campaigns.

The study of the `\textsl{O}bserved$-$\textsl{C}alculated' times of minima variation(s) (hereafter O$-$C analysis) of an EB provides valuable information for the mechanisms which form its orbital period (e.g. mass transfer, third body, magnetic influences etc). On the other hand, the `snapshot' of the binary, i.e. its light curve (hereafter LC), leads us to understand directly which physical processes are occurring (e.g. third light existence, Roche Lobe filling). The combined information from these two independent methods of analysis provides a more comprehensive view of a binary system.

Eleven systems exhibiting oscillating behaviour were selected for further observations and eight of them are studied using efficient modern techniques for LC and photometric frequency analyses. The remaining three cases QY~Aql, BO~Her, and IO~UMa, will be presented in a future work. Four of these eleven were found to have orbital period changes from O$-$C analysis.

\section[]{Observations and data reduction}
The photometric observations were carried out during September 2008 - September 2011 at the Gerostathopoulion Observatory of the University of Athens (UoA) located in Athens, Hellas, and at the Kryonerion (Kry) Astronomical Station of the Astronomical Institute of the National Observatory of Athens located at Mt.~Kyllini, Corinthia, Hellas. The instrumentation used for the observations is described in detail in Table~1.

Aperture photometry was applied to the data and differential magnitudes for all systems were obtained using the software \emph{MuniWin} v.1.1.26 \citep{HR98}. The adopted observational strategy in this survey was the same as that described in detail in the previous paper \citep{LN09}. Briefly, the observational guidelines were: i) the time span should be greater than 3~hr, ii) the filter $B$ or $V$ should be used, iii) the comparison star should be of similar magnitude and spectral type to the variable, iv) appropriate exposure times and binning modes must be used for the best possible photometric $S/N$ (signal-to-noise ratio).

In Table~2 we list: the name of the system as given in the GCVS catalogue \citep{SA12}, its apparent magnitude ($m$) and spectral type (\emph{S.T.}) as given in SIMBAD, the filters (\emph{F}) used, the number of nights (\emph{N}) of observations, the total time span (\emph{T.S.}), the standard deviation (\emph{S.D.}) of the observed points (mean value), the phase intervals (\emph{P.I.}), the dominant pulsation frequency ($f_{\rm dom}$) found and its semi-amplitude ($A_{\rm B}$) in $B$-filter (see section~6 for details), and the abbreviation for the instrumentation (\emph{In}) used for each case according to Table~1. For the majority of the systems, the observations were obtained outside the primary eclipse, since the primary component (hotter one) was the candidate $\delta$~Sct star.

The eight systems found to exhibit pulsational behaviour and whose complete LCs were obtained are listed in Table~3 along with the corresponding Comparison and Check stars.

The spectroscopic observations were obtained with the 1.3~m Ritchey-Chr\'{e}tien telescope at Skinakas Observatory, Mt.~Ida, Crete, Hellas, on 6 October 2010 for WY~Cet, UW~Cyg and on 20 and 24 September 2011 for HZ~Dra. We used a 2000$\times$800 ISA SITe CCD camera attached to a focal reducer, with a 2400~lines/mm grating and slit of 80~$\mu$m. This arrangement yielded a nominal dispersion of $\sim$0.55~\AA/pixel and wavelength coverage between 4782-5864~\AA. Data reduction was performed using the \textsl{Radial Velocity reductions} v.2.1d software \citep{NE09}. The frames were bias subtracted, a flat field correction was applied, and the sky background was removed. The spectral region was selected so as to include H$_{\beta}$ and sufficient metallic lines. Before and after each on-target observation, an arc calibration exposure (NeHeAr) was recorded.

\begin{table}
\begin{minipage}{\columnwidth}
\centering
\caption{The instrumentation setups (Telescope \& CCD) used during the photometric observations, their location ($Loc.$) and abbreviation ($Ab.$).}
\begin{tabular}{lcccc}
\hline
Telescope                              &             CCD*         &               $Loc.$          &           $Ab.$                 \\
\hline
Cassegrain - 0.4~m, $f$/8              &          ST--8XMEI       &                UoA            &             $A1$                \\
Cassegrain - 0.4~m, $f$/8              &          ST--10XME       &                UoA            &             $A2$                \\
Newt. Reflector - 0.2~m, $f$/5         &          ST--8XMEI       &                UoA            &             $A3$                \\
Newt. Reflector - 0.25~m, $f$/4.7      &          ST--8XMEI       &                UoA            &             $A4$                \\
Cassegrain - 1.2~m, $f$/13             &           AP47p          &                Kry            &             $K$                 \\
\hline
\multicolumn{5}{l}{*The CCDs are equipped with the Bessell $U,~B,~V,~R,~I$}\\
\multicolumn{5}{l}{photometric filters}
\end{tabular}
\end{minipage}
\end{table}

\begin{table*}
\centering
\caption{The total log of observations.}
\scalebox{0.94}{
\begin{tabular}{lcccccccccc}
\hline
System      &       $m$       & $S.T.$  &     $F$    &    $N$   &     $T.S.$  &   $S.D.$    &            $P.I.$      &$f_{\rm dom}$&$A_{\rm B}$&     $In$     \\
            &      (mag)      &         &            &          &    (hrs)    &    (mmag)   &                           &  (c/d)   &    (mmag) &              \\
\hline
And CP      &   11.80 ($B$)   &      A5 &     $B$    &      1   &       6     &      5.8    &           0.20-0.26       &    --    &    --     &      A2      \\
And TW      &   9.49 ($B$)    &     F0V &     $B$    &      2   &       6     &      2.2    &   0.78-0.81, 0.26-0.29    &    --    &    --     &      A2      \\
And V342    &   7.98 ($B$)    &      A3 &     $B$    &      1   &       6.5   &      1.8    &         0.90-1.00         &    --    &    --     &      A3      \\
And V363    &   9.29 ($B$)    &      A2 &     $B$    &      1   &       4     &      4.2    &          0.00-0.14        &    --    &    --     &      A3      \\
Aql QY      &    11.4 ($B$)   &     F0  &   $BVI$    &      36  &       200+  &     3.8     &           0.00-1.00       &10.655 (3)&   12.0 (1)&      A2 \& K \\
Aql V805    &   7.85 ($B$)    &      A3 &     $B$    &      1   &       4     &      1.3    &           0.82-0.88       &    --    &    --     &      A3      \\
Aql V1461   &   9.00 ($B$)    &      A0 &     $B$    &      1   &       4.5   &      3.7    &           0.56-0.67       &    --    &    --     &      A3      \\
Aqr CZ      &   11.20 ($B$)   &      A5 &     $B$    &     10   &       25+   &      2.9    &           0.00-1.00       &35.508 (2)&  3.7 (5)  &      A2      \\
Ari SZ      &   11.60 ($B$)   &      F0 &     $V$    &      1   &       3.5   &      3.6    &           0.51-0.60       &    --    &    --     &      A2      \\
Aur V417    &   7.99 ($B$)    &       A0&   $BVRI$   &      9   &       45+   &      1.5    &           0.00-1.00       &    --    &    --     &      A2      \\
Cam SS      &   10.93($B$)    &    G1III&     $B$    &      1   &       4     &      3.8    &           0.61-0.64       &    --    &    --     &      A2      \\
Cas IS      &   12.10($B$)    &       A2&     $B$    &      1   &       4     &     4.1     &           0.19-0.26       &    --    &    --     &      K       \\
Cas V364    &   11.10($B$)    &       A7&     $B$    &      1   &       4     &     3.4     &           0.15-0.26       &    --    &    --     &      A2      \\
Cas V773    &   6.32 ($B$)    &       A3&     $B$    &      1   &       4     &     2.1     &           0.45-0.59       &    --    &    --     &      A3      \\
Cas V821    &   8.37 ($B$)    &       A0&     $B$    &      1   &       4     &     1.5     &           0.60-0.68       &     ?    &    --     &      A2      \\
Cep EI      &   7.94 ($B$)    &       A5&     $B$    &      1   &       4.5   &     1.4     &           0.22-0.24       &     ?    &    --     &      A2      \\
Cep V405    &   8.95 ($B$)    &       A2&   $BVRI$   &      5   &       35    &     2.4     &           0.00-1.00       &    --    &    --     &      A2      \\
Cep WX      &   9.38 ($B$)    &       A3&     $B$    &      1   &       2.5   &     3.8     &           0.64-0.67       &    --    &    --     &      A2      \\
Cep XX      &   9.47 ($B$)    &      A7V&     $B$    &      1   &       4     &     2.5     &           0.63-0.69       & 32.07 (4)&   3.6 (4) &      A2      \\
Cet DP      &   7.01 ($B$)    &       A2&     $B$    &      1   &       4     &     1.9     &           0.88-0.92       &    --    &    --     &      A3      \\
Cyg MY      &   8.68 ($B$)    &     A2.5&     $B$    &      3   &       13    &     2.0     &   0.21-0.29, 0.71-0.76    &    --    &    --     &      A2      \\
Cyg UW      &   11.00 ($B$)   &       A5&    $BVI$   &      26  &       90+   &     2.7     &           0.00-1.00       &27.841 (2)&  1.9 (2)  &      A2      \\
Cyg V477    &   8.71 ($B$)    &      A1V&     $BV$   &      1   &       4     &     2.9     &           0.71-0.77       &    --    &    --     &      A2      \\
Cyg V959    &   11.50 ($B$)   &       A5&     $B$    &      1   &       4     &     3.4     &           0.49-0.61       &    --    &    --     &      K       \\
Cyg V2083   &   7.13 ($B$)    &       A3&     $B$    &      1   &       4     &     2.3     &           0.09-0.15       &    --    &    --     &      A3      \\
Cyg V2154   &   8.18 ($B$)    &       F0&     $B$    &      1   &       4     &     1.1     &           0.59-0.65       &    --    &    --     &      A2      \\
Cyg VW      &   10.58 ($B$)   &      A3 &     $B$    &      1   &       4     &     3.0     &           0.20-0.22       &    --    &    --     &      A2      \\
Dra HL      &   7.52 ($B$)    &       A5&   $BVRI$   &      17  &       60+   &     2.1     &           0.00-1.00       &26.914 (1)&  3.0 (2)  &      A2      \\
Dra HZ      &   8.34 ($B$)    &       A0&   $BVRI$   &      8   &       25+   &     2.7     &           0.00-1.00       &51.068 (2)&  4.0 (4)  &      A3      \\
Dra RX      &   10.84 ($B$)   &       F0&     $B$    &      2   &       11    &     5.4     &   0.38-0.45, 0.58-0.64    &    --    &   --      &      A2      \\
Her AD      &   10.02 ($B$)   &      A4V&     $B$    &      1   &       4     &     4.2     &           0.63-0.65       &    --    &   --      &      A3      \\
Her BO      &  11.6 ($B$)     &      A7 &    $BVI$   &      26  &       130   &     4.5     &           0.00-1.00       &13.430 (1)&    68 (3) &    A2 \& K   \\
Her FN      &   10.50 ($B$)   &       A8&     $B$    &      1   &       4.5   &     6.0     &           0.76-0.83       &    --    &   --      &      A2      \\
Her HS      &   8.58 ($B$)    &   B6III &    $BVI$   &      4   &       4     &     3.8     &         0.20-0.30         &    --    &   --      &      A2      \\
Her SZ      &   10.28 ($B$)   &      F0V&     $B$    &      1   &       4     &     2.8     &          0.20-0.29        &    --    &   --      &      A2      \\
Her UX      &   9.11 ($B$)    &      A0V&     $B$    &      1   &       4     &     3.6     &           0.73-0.82       &    --    &   --      &      A1      \\
Her V948    &   9.26 ($B$)    &       F2&   $BVRI$   &     10   &       35    &     3.5     &           0.00-1.00       &    --    &   --      &      A2      \\
Her V1002   &   9.14 ($B$)    &       A0&     $B$    &      1   &       4     &     3.5     &           0.13-0.23       &    --    &   --      &      A3      \\
Hya DE      &   11.00 ($B$)   &       A2&     $B$    &      1   &       4     &     3.9     &           0.28-0.32       &    --    &   --      &      A2      \\
Lac AU      &   11.50 ($B$)   &       A5&   $BVRI$   &      19  &       80+   &     2.0     &           0.00-1.00       &58.217 (1)&  5.0 (3)  &    K \& A2   \\
Lac CM      &   8.39 ($B$)    &       A3&   $BVRI$   &      10  &       40+   &     2.6     &           0.00-1.00       &    ?     &   --      &      A2      \\
Lac VX      &   10.83 ($B$)   &       F0&     $B$    &      3   &       14    &     4.5     &          0.10-0.50        &    --    &    --     &      A2      \\
Lac V364    &    8.54 ($B$)   &       A3&     $B$    &      1   &       4     &     1.9     &          0.25-0.27        &    --    &    --     &      A3      \\
Lac V398    &    8.87 ($B$)   &       A0&     $B$    &      1   &       4     &     3.1     &         0.27-0.30         &    --    &   --      &      A3      \\
Lyn CL      &   10.05 ($B$)   &       A5&    $BVI$   &     12   &      90+    &     4.0     &          0.00-1.00        &23.051 (1)&  7.3 (3)  &      A4      \\
Lyn SX      &   10.00 ($B$)   &       A2&    $BVI$   &      6   &       30    &     8.4     &   0.00-0.10, 0.48-0.65    &    --    &    --     &      A3      \\
Lyr RV      &   11.50 ($B$)   &       A5&     $B$    &      1   &       6     &     3.6     &           0.74-0.80       &    --    &    --     &      K       \\
Mon EP      &   10.50 ($V$)   &       A3&     $B$    &      1   &       4     &     3.6     &          0.20-0.34        &   --     &    --     &      A2      \\
Mon HO      &   11.40 ($B$)   &       A5&     $B$    &      1   &     5.5     &     3.9     &           0.18-0.21       &    --    &    --     &      A2      \\
Oph V391    &   11.50 ($B$)   &       A1&     $B$    &      1   &       5     &     4.5     &           0.56-0.65       &    --    &    --     &      K       \\
Oph V456    &   10.37 ($B$)   &       A2&     $B$    &      1   &       4     &     3.7     &           0.67-0.85       &    --    &    --     &      A1      \\
Ori EY      &   10.21 ($B$)   &       A7&     $B$    &      1   &       4     &     7.8     &           0.32-0.33       &   --     &    --     &      A4      \\
Ori FK      &   11.80 ($B$)   &       A2&     $B$    &      1   &       4     &     2.0     &           0.35-0.44       &    --    &    --     &      K       \\
Ori FT      &   9.35 ($B$)    &       A0&     $B$    &      1   &       4     &     2.7     &           0.72-0.77       &    --    &    --     &      A2      \\
Ori V536    &   10.50 ($B$)   &       A2&     $B$    &      1   &       3     &     3.1     &           0.46-0.50       &    --    &    --     &      A1      \\
Peg AT      &   9.21 ($B$)    &    A3.5V&    $BR$    &      8   &       30    &     3.3     &           0.00-1.00       &    --    &    --     &      A3      \\
Peg BG      &   10.50 ($B$)   &     A2V &   $BVRI$   &     15   &       102   &     5.8     &   0.00-0.29, 0.35-1.00    &25.543 (1)&  12.7 (5) &   A2 \& A4   \\
Peg DM      &   11.80 ($B$)   &       A3&    $B$     &      1   &       4     &     4.3     &           0.72-0.79       &    ?     &     --    &      A2      \\
Peg OO      &   8.53 ($B$)    &       A2&    $B$     &      1   &       4     &     2.9     &           0.79-0.83       &    --    &     --    &      A3      \\
Per RV      &   11.40 ($B$)   &       A0&    $B$     &      1   &       5     &     3.9     &           0.16-0.27       &    --    &    --     &     A2       \\
Sge UZ      &   11.40 ($B$)   &       A0&    $VR$    &      3   &       60+   &     4.9     &           0.00-1.00       &    ?     &    --     &      A2      \\
Tau EW      &   11.70 ($B$)   &       - &     $B$    &      1   &       4     &     3.1     &           0.59-0.62       &    --    &    --     &      K       \\
Tau V1149   &   8.65 ($B$)    &       A0&     $B$    &      1   &       4.5   &     2.1     &           0.63-0.67       &    --    &    --     &      A4      \\
UMa IO      &   8.44 ($B$)    &       A3&   $BVRI$   &     47   &       150+  &     1.5     &           0.00-1.00       &22.015 (2)&   6.7 (1) &   A2 \& A3   \\
UMi RT      &   11.10 ($B$)   &       F0&     $B$    &      1   &       5     &     3.7     &           0.49-0.61       &    --    &    --     &      A2      \\
Vul AW      &   10.00 ($B$)   &       F0&     $B$    &      1   &       5.5   &     4.4     &           0.59-0.88       &    --    &    --     &      A2      \\
Vul BP      &   10.17 ($B$)   &       A7&     $B$    &      1   &       3     &     2.5     &           0.24-0.30       &    --    &    --     &      A2      \\
Vul RR      &   10.15 ($B$)   &       A2&     $B$    &      2   &       6     &     2.7     &   0.16-0.18, 0.64-0.67    &    ?     &    --     &      A2      \\
\hline
\end{tabular}}
\end{table*}

\section[]{Spectroscopic analysis}

WY~Cet, UW~Cyg and a total of 23 spectroscopic standard stars ranging from A0 to G8 spectral types were observed with the same instrumental set-up. We used 900-s and 2000-s exposure times for WY~Cet and UW~Cyg, respectively. Both spectra were observed well inside the secondary eclipses (at the phase 0.506 for WY~Cet and 0.509 for UW~Cyg) when the light contribution from the secondary component is minimal, therefore the spectra practically correspond to the primaries. All spectra were calibrated and normalized to enable direct comparisons. Then, we shifted the spectra, using H$_{\beta}$ as reference, to compensate for the relative Doppler shifts of each standard. The variables' spectra were subtracted from those of each standard, deriving sums of squared residuals in each case. Such least squares sums should allow the best match between the spectra of variable and standard to be found.

\begin{table}
\centering
\caption{The photometric observations log of the eight selected EBs.}
\begin{tabular}{lll}
\hline
System          &        Comparison star   &        Check star      \\
\hline
CZ Aqr          &      GSC 6396-1024       &     GSC 6396-0872      \\
TY Cap          &      GSC 5749-2167       &     GSC 5749-1557      \\
WY Cet          &         HIP 7373         &     GSC 5279-0617      \\
UW Cyg          &      GSC 3164-0083       &     GSC 3164-0269      \\
HL Dra          &         SAO 31053        &     GSC 3913-0901      \\
HZ Dra          &         SAO 18500        &     GSC 4449-1053      \\
AU Lac          &      GSC 3610-0231       &     GSC 3610-0685      \\
CL Lyn          &      GSC 3787-0420       &     GSC 3783-0649      \\
\hline
\end{tabular}
\end{table}

Comparison between the spectra of WY~Cet and UW~Cyg and the standards yielded the primaries to be A9V and A6V type stars, respectively. Fig.~1a shows the best matching of the variables' spectra with those of the standards. The spectrum of HZ~Dra, although dominated by the primary component, could not fit sufficiently well with any of the standards. That may means that the secondary contributes in way that affects the spectrum, therefore spectral classification could not be accurate.

For the radial velocity (RV) calculations for HZ~Dra the software \emph{Broadening Functions} (BFs) v.2.4c \citep{NE09}, which is based on the method of \citet{RU02}, was used. We cropped all spectra in order to avoid the broad H$_{\beta}$ line, and we included all the sharp metallic lines between 4800-5350~\AA. Each RV value and its error was derived statistically (mean value and error) from the respective velocities resulting from BFs method by using six different standard stars of similar spectral type to the system. Due to the large brightness difference between the components of the system, we obtained measurements only for its primary. The semi-amplitude of the RV curve, $K_1$, and the systemic velocity $V_0$ were calculated by fitting a sinusoidal function to the RV points. A sample of the heliocentric RVs are given in Table~4, while the rest are given in the electronic version of the paper. The RV plot is illustrated in Fig.~1b.

\begin{table}
\centering
\caption{Sample of the heliocentric radial velocities measurements of HZ~Dra.}
\begin{tabular}{lcc}
\hline
      HJD	    &	Phase	&	  $RV_1$     \\
                &           &     (km/s)     \\
\hline
2455825.3099	&	0.2800	&	$-$36 (14)   \\
2455825.3549	&	0.3382	&	$-$35 (14)   \\
2455829.4886	&	0.6862	&	28 (12)      \\
\hline
\end{tabular}
\end{table}

\begin{figure}
\begin{tabular}{cc}
\includegraphics[width=7.2cm]{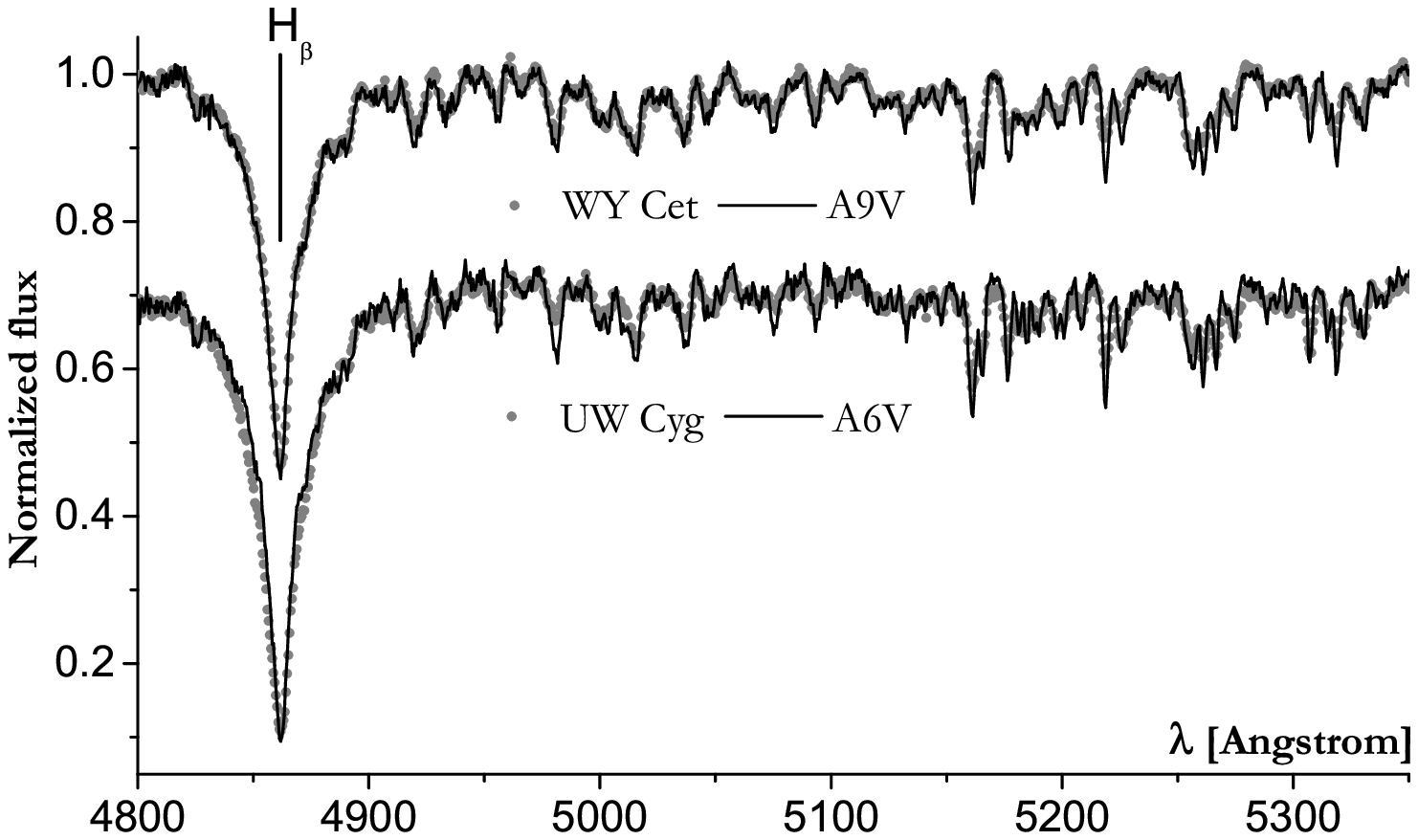}&(a)\\
\includegraphics[width=7.2cm]{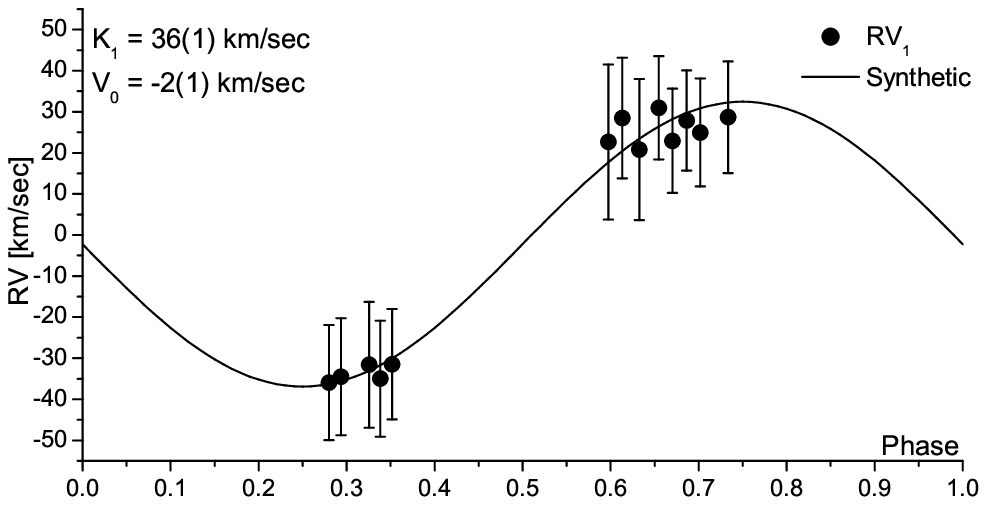}&(b)\\
\end{tabular}
\caption{(a): The comparison spectra of WY~Cet (upper) and UW~Cyg (lower) and the standard stars A9V (HIP~11678) and A6V (HIP~21589), respectively. (b): Synthetic (solid line) and observed (points) radial velocities of the primary component of HZ~Dra. The radial velocity amplitude $K_1$ and the systemic velocity $V_0$ are also indicated.}
\label{fig1}
\end{figure}

\section[]{Light curve analysis and absolute parameters derivation}

Complete LCs of each system were analysed using \emph{PHOEBE} v.0.29d software \citep{PZ05} that follows the 2003 version of the Wilson-Devinney (WD) code \citep{WD71,WI79,WI90}. In the cases where multicolour photometry was available, the LCs were analysed simultaneously. For HL~Dra and HZ~Dra the radial velocities of the primary component (\citet{PR06} and present paper) were included in the analysis. In the absence of spectroscopic mass ratios, the `$q$-search' method using a step of 0.1 was trialled in Modes 2 (detached system) and 5 (conventional semi-detached system) to find `photometric' estimates for the mass ratio $q_{\rm ph}$. This value was then set as initial input and treated as a free parameter in the subsequent analysis. The temperatures of the primaries were assigned values according to their spectral types using the correlations of \citet{CO00} and were kept fixed, while the temperatures of the secondaries $T_2$ were adjusted. The values of bolometric albedos $A_1$ and $A_2$, and gravity darkening coefficients, $g_1$ and $g_2$, were set as $A$=1 and $g$=1 for radiative \citep{RU69,VZ24} and $A$=0.5 and $g$=0.32 for convective atmospheres \citep{RU69,LU67}. Synchronous rotation was assumed, so the synchronization parameters $F_1$ and $F_2$ were set as 1. The linear limb darkening coefficients, $x_1$ and $x_2$, were taken from the tables of \citet{VH93}; the dimensionless potentials $\Omega_{1}$ and $\Omega_{2}$, the fractional luminosity of the primary component $L_{1}$ and the inclination $i$ of the system's orbit were set as adjustable. Since the O$-$C diagrams (see next section) of CZ~Aqr, TY~Cap, WY~Cet and UW~Cyg suggested possible existence of other components, the third light parameter $l_3$ was also adjusted. Best-fit models and observed LCs of the systems are presented in Fig.~2 with corresponding parameters in Table~5.

\begin{landscape}
\begin{table}
\centering
\caption{Light curve solution and absolute parameters of the components (P=Primary, S=Secondary).}
\scalebox{0.85}{
\begin{tabular}{l cccc cccc cccc cccc}
\hline
System:&\multicolumn{2}{c}{CZ Aqr}  & \multicolumn{2}{c}{TY Cap}&\multicolumn{2}{c}{WY Cet}  & \multicolumn{2}{c}{UW Cyg}&\multicolumn{2}{c}{HL Dra}  & \multicolumn{2}{c}{HZ Dra}&\multicolumn{2}{c}{AU Lac}  & \multicolumn{2}{c}{CL Lyn}\\
\hline
                                                            \multicolumn{16}{c}{\textsl{Light curve parameters}}\\
\hline
Mode	&	\multicolumn{2}{c}{Semidetached}	&	\multicolumn{2}{c}{Semidetached}	&	\multicolumn{2}{c}{Semidetached}	&	\multicolumn{2}{c}{Semidetached}	&	\multicolumn{2}{c}{Semidetached}	&	 \multicolumn{2}{c}{Detached}	&	\multicolumn{2}{c}{Semidetached}	&	\multicolumn{2}{c}{Semidetached}	 \\
$i~(\degr$)	&	\multicolumn{2}{c}{89.7 (1)}	&	\multicolumn{2}{c}{80.4 (2)}	&	\multicolumn{2}{c}{81.8 (1)}	&	\multicolumn{2}{c}{87.1 (1)}	&	 \multicolumn{2}{c}{66.5 (1)}	&	\multicolumn{2}{c}{72.0 (3)}	 &	 \multicolumn{2}{c}{83.0 (1)}	&	\multicolumn{2}{c}{78.7 (1)}	\\
$q~(m_{2}/m_{1}$)	&	\multicolumn{2}{c}{0.49 (1)}	&	\multicolumn{2}{c}{0.52 (1)}	&	\multicolumn{2}{c}{0.26 (1)}	&	\multicolumn{2}{c}{0.14 (1)}	 &	\multicolumn{2}{c}{0.37 (1)}	&	 \multicolumn{2}{c}{0.12 (4)}	&	\multicolumn{2}{c}{0.30 (1)}	&	\multicolumn{2}{c}{0.19 (2)}	\\
\hline
\textsl{Component:}&	P	&	S	&	P	&	S	&	P	&	S	&	P	&	S	&	P	&	S	&	P	&	S	&	P	&	S	&	P	&	S	\\
\hline
$T$ (K)	&	8200$^1$	&	5650 (12)	&	8200$^1$	&	4194 (30)	&	7500$^2$	&	4347 (7)	&	8000$^2$	&	4347 (4)	&	8200$^3$	&	5074 (8)	&	9800$^4$	&	5015 (68)	&	8200$^5$	&	3784 (15)	 &	8200$^6$	&	4948 (14)	\\
$\Omega$	&	3.44 (1)	&	2.86	&	3.80 (2)	&	2.93	&	4.18 (1)	&	2.40	&	5.89 (1)	&	2.07	&	2.93 (1)	&	2.63	&	 2.48 (1)	&	2.28 (2)	&	4.51 (2)	&	2.46	&	3.36 (1)	 &	 2.21	\\
$x_{\rm B}$	&	0.584	&	0.760	&	0.614	&	1.004	&	0.701	&	0.968	&	0.596	&	0.971	&	0.596	&	0.853	&	0.491	&	0.862	&	 0.540	 &	0.831	&	0.588	&	0.874	\\
$x_{\rm V}$	&	--	&	--	&	0.510	&	0.850	&	0.570	&	0.822	&	0.517	&	0.820	&	0.509	&	0.708	&	0.418	&	0.721	&	0.474	&	 0.747	&	0.505	&	0.729	\\
$x_{\rm R}$	&	--	&	--	&	0.424	&	0.738	&	0.476	&	0.708	&	--	&	--	&	0.426	&	0.611	&	0.353	&	0.622	&	0.402	&	0.691	 &	 --	&	--	\\
$x_{\rm I}$	&	--	&	--	&	0.331	&	0.606	&	0.367	&	0.586	&	0.351	&	0.591	&	0.338	&	0.513	&	0.281	&	0.523	&	0.327	&	 0.564	&	0.340	&	0.527	\\
($L/L_{\rm T})_{\rm B}$	&	0.878 (2)	&	0.122 (1)	&	0.953 (4)	&	0.018 (1)	&	0.946 (2)	&	0.044 (1)	&	0.945 (2)	&	0.051 (1)	&	0.956 (1)	&	 0.044 (1)	&	0.995 (1)	&	0.005 (1)	&	 0.987 (1)	 &	 0.013 (1)	&	0.959 (6)	&	0.042 (2)	\\
($L/L_{\rm T})_{\rm V}$	&	--	&	--	&	0.938 (6)	&	0.037 (1)	&	0.915 (2)	&	0.076 (1)	&	0.902 (3)	&	0.093 (2)	&	0.932 (1)	&	0.068 (1)	&	 0.992 (1)	&	0.008 (1)	&	0.969 (1)	&	 0.031 (1)	 &	 0.934 (8)	&	0.066 (4)	\\
($L/L_{\rm T})_{\rm R}$	&	--	&	--	&	0.914 (8)	&	0.056 (1)	&	0.883 (2)	&	0.106 (2)	&	--	&	--	&	0.912 (1)	&	0.088 (1)	&	0.988 (1)	&	 0.012 (2)	&	0.948 (2)	&	0.052 (2)	&	--	 &	--	 \\
($L/L_{\rm T})_{\rm I}$	&	--	&	--	&	0.887 (10)	&	0.084 (2)	&	0.843 (3)	&	0.144 (3)	&	0.802 (4)	&	0.180 (1)	&	0.888 (1)	&	0.112 (1)	&	 0.984 (1)	&	0.016 (4)	&	0.916 (2)	&	 0.084 (2)	 &	 0.887 (10)	&	0.113 (1)	\\
($L_3/L_{\rm T})_{\rm B}$	&	\multicolumn{2}{c}{--}	&	\multicolumn{2}{c}{0.029 (3)}	&	\multicolumn{2}{c}{0.010 (1)}	&	\multicolumn{2}{c}{0.004 (1)}	 &	 \multicolumn{2}{c}{--}	&	\multicolumn{2}{c}{--}	&	 \multicolumn{2}{c}{--}	&	\multicolumn{2}{c}{--}	\\
($L_3/L_{\rm T})_{\rm V}$	&	\multicolumn{2}{c}{--}	&	\multicolumn{2}{c}{0.025 (4)}	&	\multicolumn{2}{c}{0.009 (2)}	&	\multicolumn{2}{c}{0.005 (1)}	 &	 \multicolumn{2}{c}{--}	&	\multicolumn{2}{c}{--}	&	 \multicolumn{2}{c}{--}	&	\multicolumn{2}{c}{--}	\\
($L_3/L_{\rm T})_{\rm R}$	&	\multicolumn{2}{c}{--}	&	\multicolumn{2}{c}{0.030 (6)}	&	\multicolumn{2}{c}{0.010 (2)}	&	\multicolumn{2}{c}{--}	&	 \multicolumn{2}{c}{--}	&	\multicolumn{2}{c}{--}	&	 \multicolumn{2}{c}{--}	 &	\multicolumn{2}{c}{--}	\\
($L_3/L_{\rm T})_{\rm I}$	&	\multicolumn{2}{c}{--}	&	\multicolumn{2}{c}{0.029 (8)}	&	\multicolumn{2}{c}{0.013 (2)}	&	\multicolumn{2}{c}{0.018 (1)}	 &	 \multicolumn{2}{c}{--}	&	\multicolumn{2}{c}{--}	&	 \multicolumn{2}{c}{--}	&	\multicolumn{2}{c}{--}	\\
\hline
                                                            \multicolumn{16}{c}{\textsl{Absolute parameters}}\\
\hline
$M~(M_{\sun}$)	&	2.00$^a$	&	0.98 (1)	&	2.00$^a$	&	1.05 (1)	&	1.70$^a$	&	0.44 (1)	&	1.90$^a$	&	0.26 (1)	&	2.5 (2)	&	 0.9 (1)	&	3.0 (3)	&	0.4 (1)	&	2.0$^a$	&	0.60 (1)	 &	 2.0$^a$	&	0.38 (3)	\\
$R~(R_{\sun}$)	&	1.9 (1)	&	1.8 (1)	&	2.5 (1)	&	2.5 (1)	&	2.2 (1)	&	2.3 (1)	&	2.2 (1)	&	2.9 (1)	&	2.5 (4)	&	1.8 (3)	&	2.3 (1)	&	0.8 (1)	&	 1.8 (1)	&	2.1 (1)	&	2.5 (1)	&	1.9 (1)	\\
$L~(L_{\sun}$)	&	15.3 (9)	&	2.9 (2)	&	24.3 (8)	&	1.8 (2)	&	14.0 (9)	&	1.7 (1)	&	18.0 (9)	&	2.6 (1)	&	24.3 (7)	&	1.9 (1)	&	 45 (3)	&	0.4 (2)	&	12.6 (7)	&	0.8 (1)	&	25.2 (9)	 &	 2.0 (7)	\\
$M_{\rm bol}$ (mag)	&	1.8 (6)	&	3.6 (6)	&	1.3 (1)	&	4.1 (1)	&	1.9 (8)	&	4.2 (8)	&	1.6 (4)	&	3.7 (6)	&	1.3 (2)	&	4.1 (2)	&	0.6 (4)	&	5.9 (4)	&	 2.0 (6)	&	5.0 (7)	&	1.2 (9)	&	4.0 (8)	\\
\textsl{a}~($R_{\sun}$)	&	1.9 (2)	&	3.8 (1)	&	2.7 (3)	&	5.2 (1)	&	1.8 (3)	&	6.9 (1)	&	1.5 (2)	&	11.2 (1)	&	1.7 (3)	&	4.4 (1)	&	0.6 (1)	&	 4.7 (2)	&	1.7 (2)	&	5.7 (1)	&	1.3 (3)	&	6.6 (1)	 \\
\hline
\multicolumn{17}{l}{$^1$\citet{PE97}, $^2$present paper (section~3), $^3$\citet{PR06}, $^4$\citet{WR03}, $^5$\citet{BU04}, $^6$\citet{AD01}, $^a$assumed, $L_{\rm T}=L_1+L_2+L_3$}
\end{tabular}}
\end{table}

\begin{figure}
\begin{tabular}{cccc}
\includegraphics[width=5.7cm]{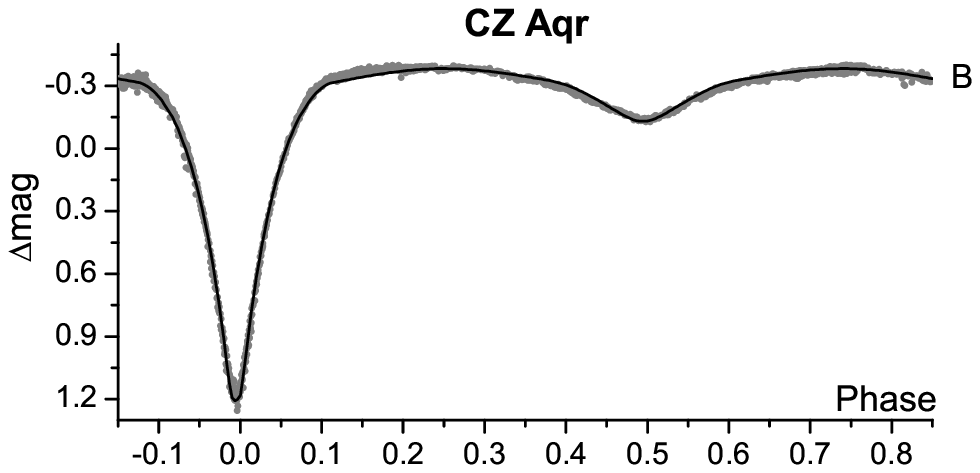}&\includegraphics[width=5.7cm]{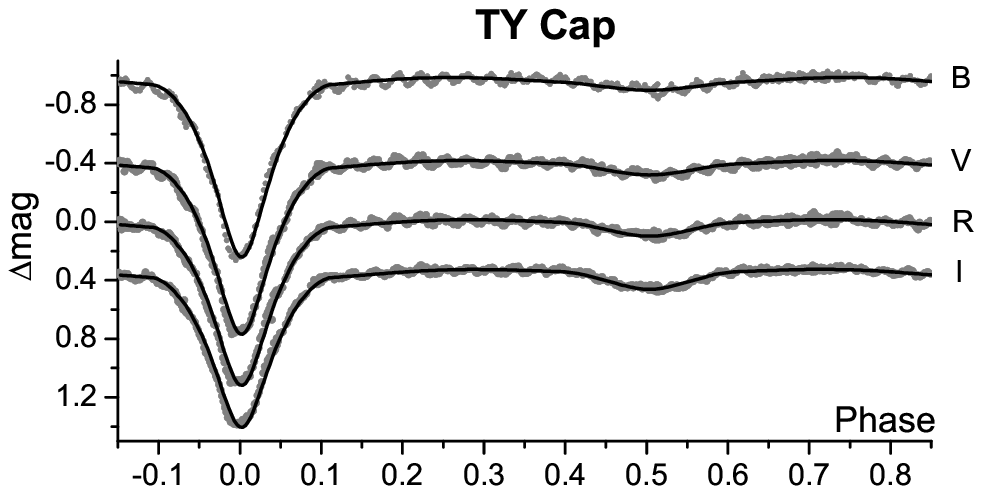}&\includegraphics[width=5.7cm]{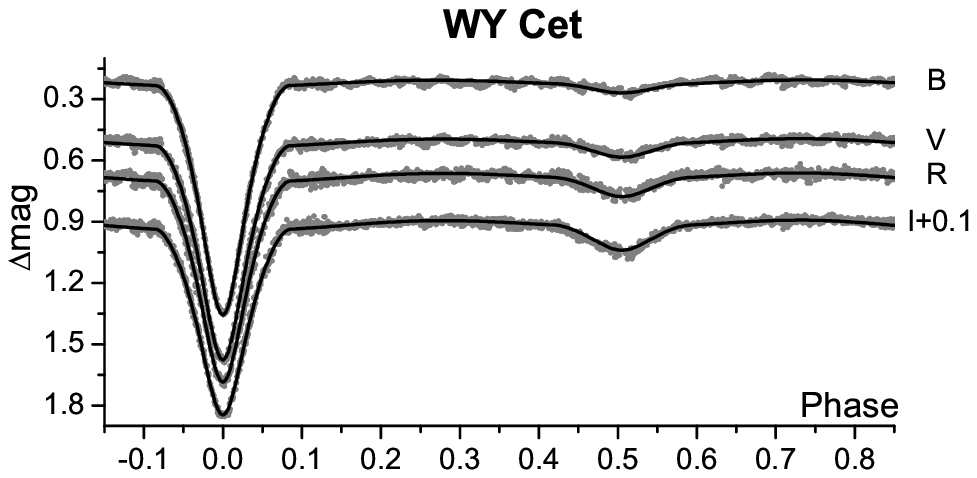}&\includegraphics[width=5.7cm]{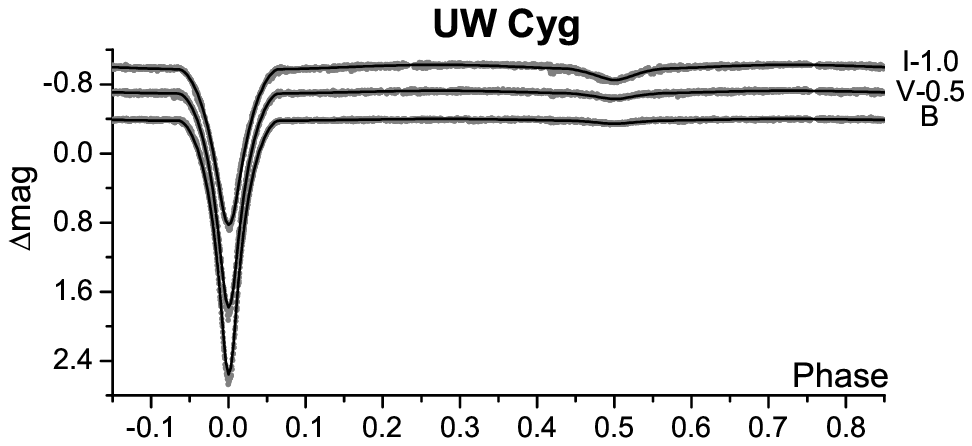}\\
\includegraphics[width=5.7cm]{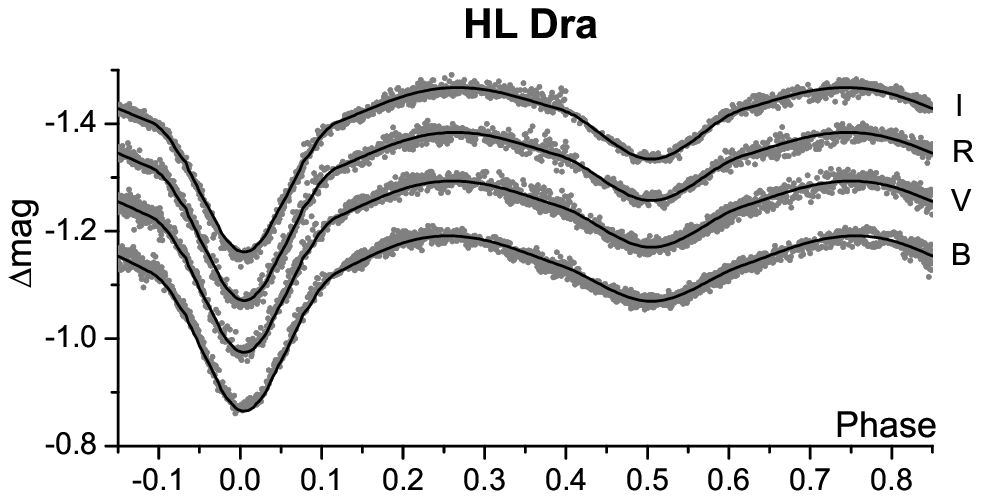}&\includegraphics[width=5.7cm]{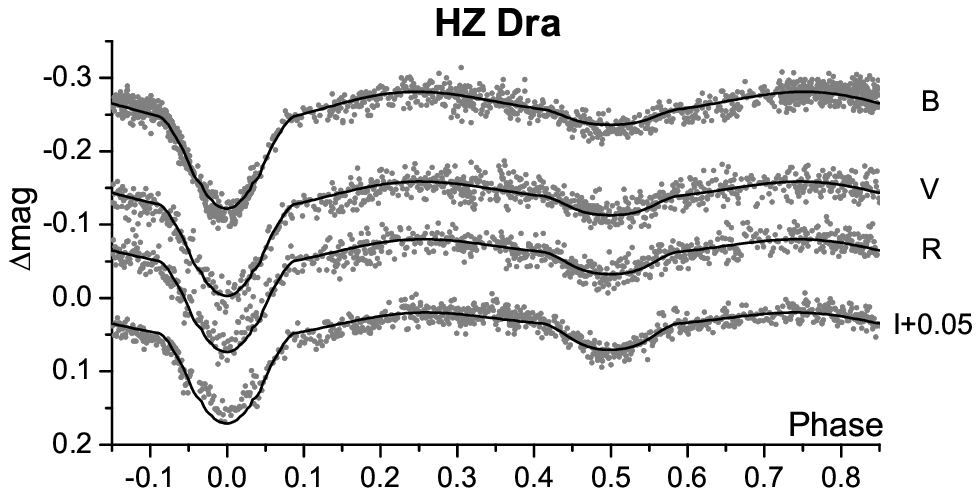}&\includegraphics[width=5.7cm]{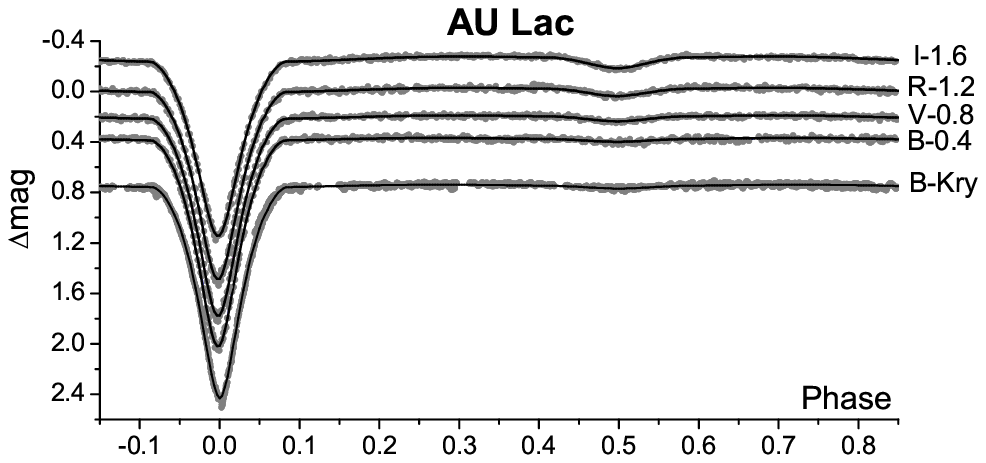}&\includegraphics[width=5.7cm]{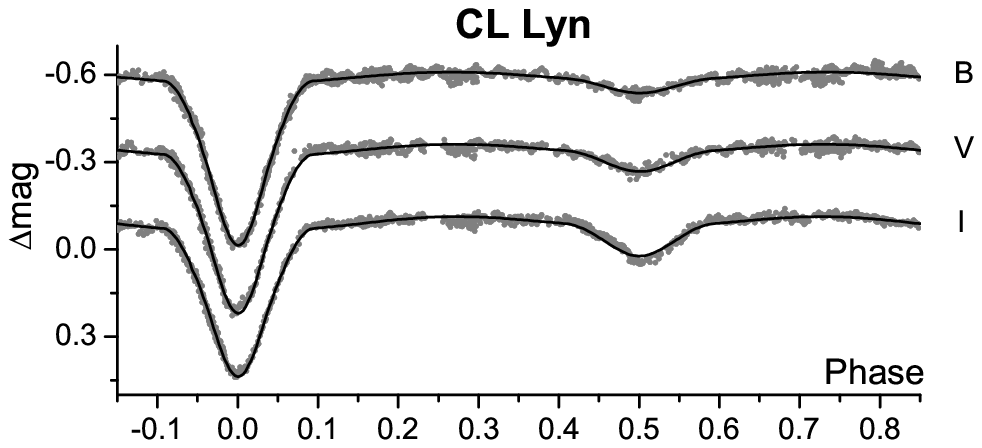}\\
\end{tabular}
\caption{Synthetic (solid lines) and observed (grey points) light curves of the systems.}
\label{fig2}
\end{figure}
\end{landscape}

Although no radial velocity curves for these systems were published as yet, except for the primaries of HL~Dra and HZ~Dra, we can form fair estimates of their absolute parameters (see Table~5). The masses of the primaries were assumed from their spectral type, while those of the secondaries followed from the adopted mass ratios. The semi-major axes (\textsl{a}), used to calculate mean radii, follow from Kepler's law. In Fig.~3 the position of the systems' components in the Mass-Radius ($M-R$) diagram is presented. The theoretical lines for Zero Age Main Sequence (ZAMS) and Terminal Age Main Sequence (TAMS) were taken from \citet{NM03}.

\begin{figure}
\includegraphics[width=8cm]{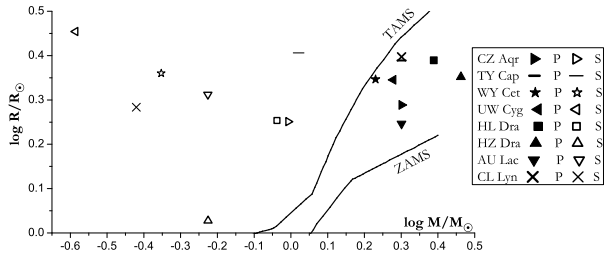}
\caption{The location of the systems' components ($P$ for primary and $S$ for the secondary) in the $M-R$ diagram. The primaries and the secondaries belonging to near contact systems are indicated with grey symbols for comparison.}
\label{fig3}
\end{figure}

\section{Orbital period analyses}

\begin{table}
\centering
\caption{O$-$C diagram analyses results.}
\scalebox{0.85}{
\begin{tabular}{lccc}
\hline
System                      &        CZ Aqr         &        \multicolumn{2}{c}{UW Cyg}       \\
\hline
$JD_0$ (HJD)                &     2443371.449 (2)   &   \multicolumn{2}{c}{2443690.082 (3)}   \\
$P$ (d)                     &       0.862752 (3)    &   \multicolumn{2}{c}{3.450759 (1)}      \\
\hline
                            &       3$^{rd}~body$   &      3$^{rd}~body$   &   4$^{th}~body$  \\
\hline
$T_0$ (HJD)                 &        2455535 (8000) &      2435563 (1000)  &     2436101 (878)\\
$\omega~(\degr$)            &         166 (22)      &        213 (13)      &       78 (34)    \\
$A$ (d)                     &         0.026 (4)     &       0.061 (2)      &       0.013 (1)  \\
$P'$ (yr)                   &        103 (5)        &        82 (2)        &       30 (1)     \\
$e$                         &        0.3 (2)        &       0.2 (1)        &        0.4 (2)   \\
$f(M)~(M_{\sun}$)           &        0.010 (2)      &       0.184 (1)      &      0.013 (1)   \\
$M_{\rm min}~(M_{\sun}$)    &        0.49 (6)       &       1.30 (1)       &      0.59 (1)    \\
\hline
$\Sigma Wres^{2}$           &         0.014         &         \multicolumn{2}{c}{0.061}       \\
\hline
$\Delta Q$~(g~cm$^2$)       & 1.4 (2)$\times10^{50}$&3.5 (1)$\times10^{50}$&2.0 (5)$\times10^{50}$\\
\hline
System                      &           TY Cap      &            WY Cet                       \\
\hline
$JD_0$ (HJD)                &   2440523.095 (2)     &   2431553.889 (9)    &                  \\
$P$ (d)                     &      1.423458 (2)     &     1.939689 (1)     &                  \\
$C_{2}$~(d/cycle)           &7 (2)$\times10^{-11}$  &1.958 (1)$\times10^{-9}$&                \\
$\dot{P}$~(d/yr)            &3.9 (1)$\times10^{-8}$ &7.37 (1)$\times10^{-7}$&                 \\
$\dot{M}~(M_{\sun}$/yr)     &2.05 (6)$\times10^{-8}$&7.52 (1)$\times10^{-8}$&                 \\
\hline
                            &     3$^{rd}~body$     &    3$^{rd}~body$     &                  \\
\hline
$T_0$ (HJD)                 &       2465187 (4000)  &     2420356 (5000)   &                  \\
$\omega~(\degr$)            &        99 (41)        &         60 (6)       &                  \\
$A$ (d)                     &        0.029 (2)      &       0.052 (1)      &                  \\
$P'$ (yr)                   &        80 (3)         &          98 (6)      &                  \\
$e$                         &        0.2 (1)        &        0.3 (1)       &                  \\
$f(M)~(M_{\sun}$)           &        0.019 (7)      &        0.078 (9)     &                  \\
$M_{\rm min}~(M_{\sun}$)    &        0.64 (6)       &        0.89 (7)      &                  \\
\hline
$\Sigma Wres^{2}$           &         0.010         &         0.189        &                  \\
\hline
$\Delta Q$ (g~cm$^2$)       &4.4 (3)$\times10^{50}$ &3.2 (1)$\times10^{50}$&                  \\
\hline
\end{tabular}}
\end{table}

Computation of the orbital parameters of additional bodies causing cyclic period changes ($\textsl{LI}ght-\textsl{T}ime$ $\textsl{E}ffect$) is an inverse problem for several parameters, namely the period $P'$ and eccentricity $e$ of each additional body, HJD of the periastron passage $T_0$, semi-amplitude $A$ of each LITE and argument of periastron $\omega$. The mass transfer/loss rate should be accompanied by period changes that can be computed directly by using the coefficient $C_{2}$ of the parabola which fits the O$-$C points in such cases (cf. \citealt{HI01}).

The ephemeris of the EB ($JD_0$ and $P$ for the linear form and $C_{2}$ for the quadratic) should be calculated together with the parameters of the LITE. From the LITE parameters, the mass function $f(M)$ and the minimal mass $M_{\rm min}$ (assuming a coplanar orbit with the EB) of each additional body can be estimated.

A cool component of an EB is candidate for magnetic activity, hence the potential variation of its quadrupole moment $\Delta Q$ should be checked against cyclic period modulation as suggested by \citet{AP92}. For the $\Delta Q$ calculation the relations of \citet{LR02} and \citet{RO00} and the absolute parameters of the cooler component of each system (see Table~5) were used. According to the criterion of \citet{LR02} the magnetic activity of a component implies the cyclic period changes of the EB if its quadrupole moment lies between the range $10^{50}<\Delta Q<10^{51}$~g~cm$^2$.

Only CZ~Aqr, TY~Cap, WY~Cet and UW~Cyg are systems showing peculiar period modulation. Their O$-$C diagrams were analysed using the least squares method with statistical weights in a $MATLAB$-code designed by P.Z. \citep{ZA09}. Weights were set at $w=1$ for visual, 5 for photographic and 10 for CCD and photoelectric observations. Times of minima were taken from the literature (168 for CZ~Aqr, 107 for TY~Cap, 75 for WY~Cet and 193 for UW~Cyg). The input ephemerides used to compute the O$-$C points of the compiled data were taken from \citet{KR01}. In Fig.~4, full circles represent times of primary minima and open circles those of the secondary minima, where the bigger the symbol, the bigger the weight assigned. A LITE function associated with a cyclic distribution of O$-$C points and a parabola in accordance with possible mass transfer (see Table~5-geometrical status) were the respective fitting functions.

\begin{figure}
\centering
\includegraphics[width=8cm]{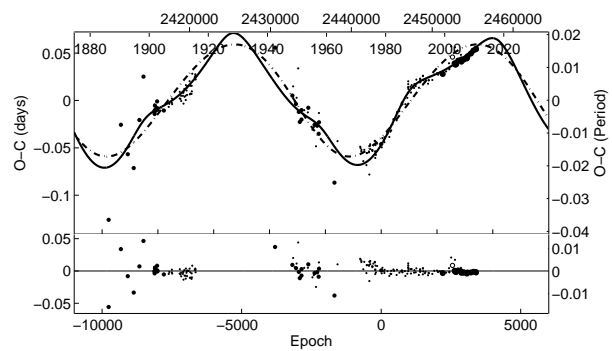}
\caption{The fitting (solid line) on the O$-$C points (upper part) and the residuals (lower part) of the total solution for UW~Cyg. Dashed dotted line corresponds to the first LITE function.}
\label{fig4}
\end{figure}

For TY~Cap and WY~Cet the adopted fitting functions describe the data very well. For CZ~Aqr and UW~Cyg the parabolic term resulted in a value very close to zero, therefore it was excluded from the final solutions. In addition, for UW~Cyg a second periodicity in the residuals was detected and a second LITE function was trialled. In Fig.~4 the O$-$C fitting for UW~Cyg (for the rest systems see electronic version) is shown. The solution parameters are listed in Table~6.

\section{Search for Pulsations}

To prepare for frequency analysis, eclipses and proximity effects were excluded by subtracting the eclipsing binary model from the observed data. Frequency analysis was performed with the software \emph{PERIOD04} v.1.2 \citep{LB05} that is based on classical Fourier analysis. Given that typical frequencies for $\delta$~Sct stars range between 3-80~c/d \citep{BR00,SO06b}, the analysis was made for this range. After the first frequency computation the residuals were subsequently pre-whitened for the next one. For cases showing pulsations, the dominant frequency is given in Table~2, where ambiguous results have a question mark (?).

Frequency analysis results for the eight selected EBs (see Table~3) are given in Table~7, where we list for each filter data set: frequency values $f$, $l$-degrees, semi-amplitudes $A$, phases $\Phi$ and $S/N$. The pulsation mode was identified from the software \emph{FAMIAS} v.1.01 \citep{ZI08}, based on $\delta$~Scuti $MAD$ models \citep{MD07} for those cases where more than one filter used. The frequency search of CZ~Aqr, UW~Cyg, HZ~Dra and AU~Lac was based only in $B$ filter data, therefore $l$-degrees were not determined for these systems. However, for some cases, frequencies having $S/N<$4 in $R$ and $I$ also appeared. These frequencies were also detected in $B$ and $V$ data, with $S/N>$4, hence we concluded that despite their being below the programme's critical limit, they are probably real. Amplitude spectra, spectral window plot and Fourier fit on the longest data set are plotted for the case of AU~Lac in Fig.~5, while for the rest systems are given in the electronic version of the paper.

For CZ~Aqr one oscillation frequency, with an amplitude very near the $4-\sigma$ significance limit, was found. For UW~Cyg and HZ~Dra the analysis was performed only on the $B$-data, while data from the other filters were neglected due to their high noise. We detected two pulsation frequencies for UW~Cyg and one for HZ~Dra, with their respective amplitudes marginally overriding the significance limit. For these three systems future observations with higher accuracy are needed.

Due to the faintness of AU~Lac ($m_{\rm B}$=11.5~mag) the $BVRI$ data obtained with the $A2$ instrument (see Table~1) did not show any clear evidence of pulsations, they were therefore excluded from the search. Although the LC modelling of AU~Lac was based on both data sets ($BVRI$ with the 40~cm and $B$ with the 1.2~m telescope), the analysis was performed only to the $B$-LC residuals of the data taken with the $K$-telescope (see Table~1). From this data set, two pulsation frequencies, with the most dominant at $\sim$58.22~c/d, were revealed.

\begin{figure}
\centering
\begin{tabular}{cl}
\includegraphics[width=6.5cm]{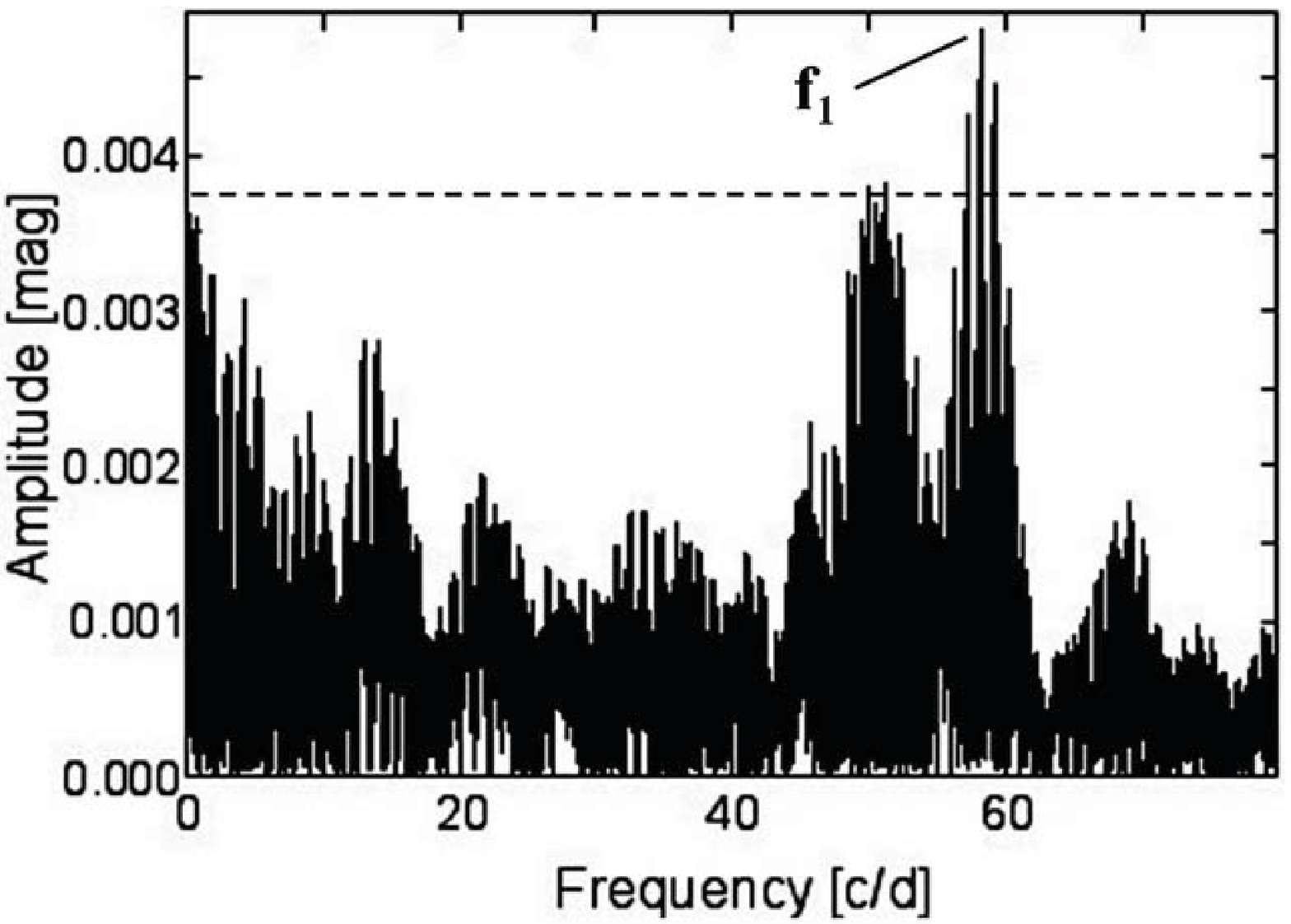}&(a)\\
\includegraphics[width=6.5cm]{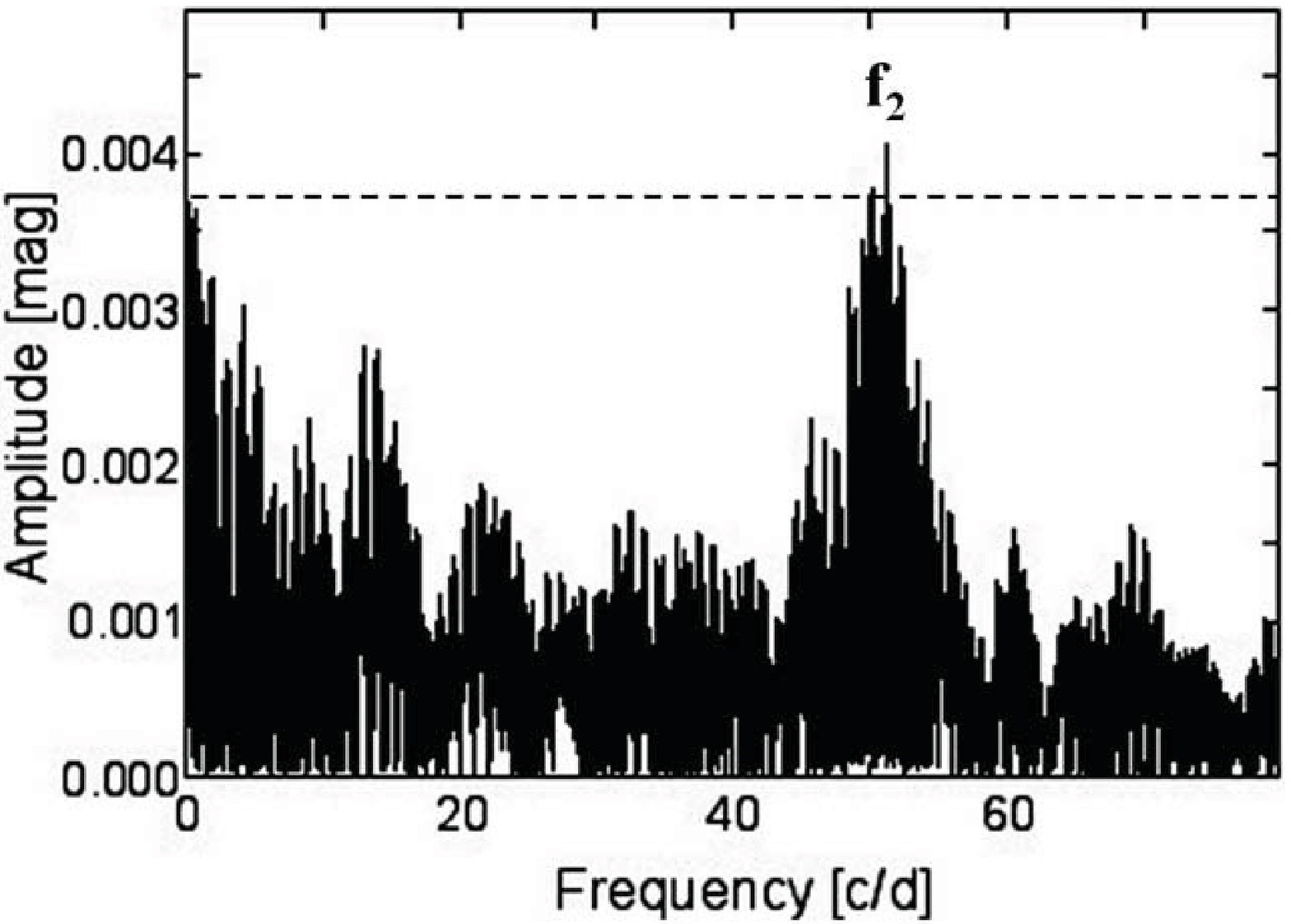}&(b)\\
\includegraphics[width=6.5cm]{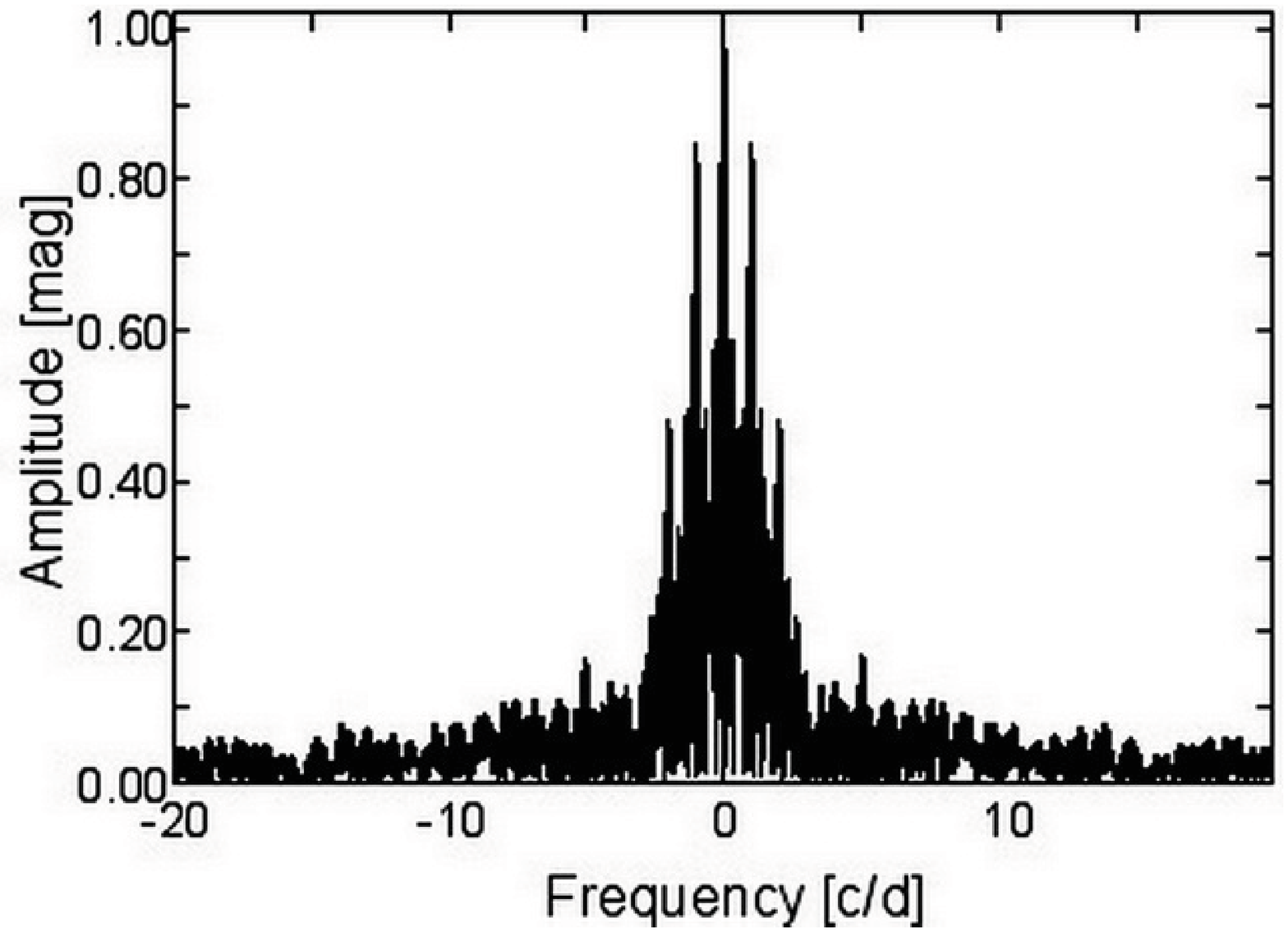}&(c)\\
\includegraphics[width=6.5cm]{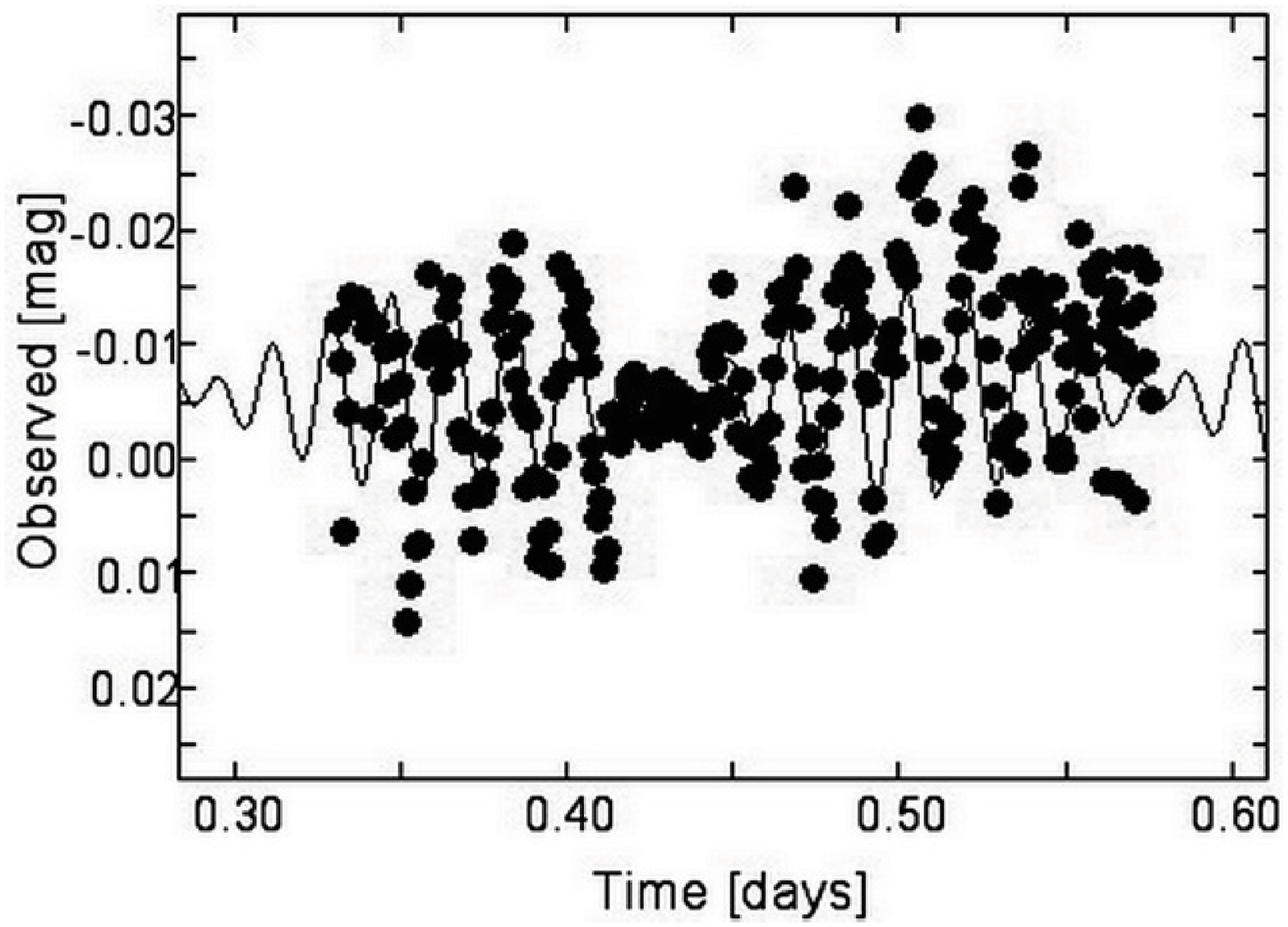}&(d)
\end{tabular}
\caption{(a), (b): Amplitude spectra, where the detected frequencies and the significance level are indicated, (c) spectral window plot, and (d) Fourier fit on the longest data set for AU~Lac.}
\label{fig5}
\end{figure}

The pulsation analysis for HL~Dra, TY~Cap and WY~Cet resulted in three, two and one frequencies, respectively. The $l$-degrees for HL~Dra were not found among $MAD$ models for the mass given in Table~5, therefore a mass value of 2~$M_{{\sun}}$ was adopted for the mode identification. Future spectroscopy observations may resolve this discrepancy.

Three frequencies in $B$ and $V$ data were traced for CL~Lyn. In the $I$-observations only one dominant frequency $f_{1}$ was found to have a $S/N>4$. The ratio $f_{1}/f_{3}\sim0.75$ is typical for the radial fundamental and first overtone modes.

\begin{landscape}
\begin{table}
\caption{Frequency analysis results for all systems.}
\centering
\scalebox{0.95}{
\begin{tabular}{lc|cccc|cccc|cccc|cccc}
\hline
        &                                                               \multicolumn{17}{c}{TY Cap}                                                           \\
\hline
        &   &   \multicolumn{4}{c}{$B$-filter}   &    \multicolumn{4}{c}{$V$-filter}  &    \multicolumn{4}{c}{$R$-filter} &   \multicolumn{4}{c}{$I$-filter}  \\
\hline
$No$    &$l$&   $F$    &  $A$   &   $\Phi$ &$S/N$&    $F$   &  $A$   &  $\Phi$  &$S/N$&   $F$    &   $A$ &  $\Phi$  &$S/N$&   $F$    & $A$   &  $\Phi$  &$S/N$\\
        &   &  (c/d)   &(mmag)  &($\degr$) &     &   (c/d)  &(mmag)  & ($\degr$)&     &  (c/d)   &(mmag) & ($\degr$)&     &  (c/d]   &(mmag) & ($\degr$)&     \\
\hline
$f_1$	& 2 &24.222 (1)&18.5 (7)&	143 (2)& 8.5 &24.224 (1)&15.6 (7)&	142 (2)	&7.2  &24.225 (2)&1.2 (6)&145 (3)   &8.6  &24.226 (1)&9.4 (5)&	142 (3) &10.1 \\
$f_2$	& - &24.590 (1)&11.3 (7)&	93 (4) & 5.0 &24.596 (1)&7.1 (7) &	85 (5)	&3.3  &24.585 (1)&6.5 (6)&114 (5)   &5.5  &24.597 (1)&5.0 (5)&	76 (5)  &5.3  \\
\hline
        &                                                               \multicolumn{17}{c}{WY Cet}                                                           \\
\hline
        &   &   \multicolumn{4}{c}{$B$-filter}   &     \multicolumn{4}{c}{$V$-filter} &  \multicolumn{4}{c}{$R$-filter}   &   \multicolumn{4}{c}{$I$-filter}  \\
\hline
$No$    &$l$&   $F$    &   $A$ &   $\Phi$ &$S/N$ &    $F$   & $A$    &  $\Phi$  &$S/N$&   $F$    &  $A$  &  $\Phi$  &$S/N$&   $F$    &  $A$  &  $\Phi$  &$S/N$\\
        &   &  (c/d)   &(mmag) &($\degr$) &      &   (c/d)  &(mmag)  & ($\degr$)&     &  (c/d)   &(mmag) & ($\degr$)&     &  (c/d]   &(mmag) & ($\degr$)&     \\
\hline
$f_1$&0 or 1&13.211 (1)&7.7 (3)&  56 (2)  & 6.0	 &13.212 (1)&5.2 (4) &   55 (5)	& 4.6 &13.210 (1)&4.5 (5)&   57 (6) & 4.8 &13.213 (1)&3.7 (4)&	40 (7)	&4.2  \\
\hline
        &                                                               \multicolumn{17}{c}{HL Dra}                                                           \\
\hline
        &   &   \multicolumn{4}{c}{$B$-filter}   &   \multicolumn{4}{c}{$V$-filter}   &  \multicolumn{4}{c}{$R$-filter}    & \multicolumn{4}{c}{$I$-filter}   \\
\hline
$No$    &$l$&   $F$    &   $A$  &   $\Phi$ &$S/N$&    $F$   &  $A$   &  $\Phi$  &$S/N$&   $F$    &  $A$  &  $\Phi$  &$S/N$&   $F$    & $A$   &  $\Phi$  &$S/N$\\
        &   &  (c/d)   &(mmag)  &($\degr$) &     &   (c/d)  &(mmag)  & ($\degr$)&     &  (c/d)   &(mmag) & ($\degr$)&     &  (c/d]   &(mmag) & ($\degr$)&     \\
\hline
$f_1$	&2  &26.914 (1)&3.0 (2)	&248 (4)   &5.0	 &26.914 (1)&2.8 (3) &	245 (11)&5.0  &26.911 (1)&1.9 (3)&261 (10)  &2.9  &26.916 (1)&1.9 (3)&	224 (8)	&4.0  \\
$f_2$&1 or 2&31.072 (1)&2.7 (2)	&221 (5)   &4.8	 &31.072 (1)&1.5 (3) &	225 (7)	&2.7  &31.071 (1)&1.5 (3)&228 (13)  &3.2  &31.072 (1)&2.0 (3)&	239 (7)	&4.0  \\
$f_3$	&-  &23.795 (1)&2.7 (2)	&319 (5)   &5.1	 &23.795 (1)&2.1 (3) &	318 (12)&4.5  &23.793 (1)&1.5 (3)&317 (13)  &3.0  &23.796 (1)&1.4 (3)&	308 (10)&2.8  \\
\hline
        &                                                               \multicolumn{17}{c}{CL Lyn}                                                           \\
\hline
        &   &   \multicolumn{4}{c}{$B$-filter}   &   \multicolumn{4}{c}{$V$-filter}   &   \multicolumn{4}{c}{$R$-filter}  &  \multicolumn{4}{c}{$I$-filter}   \\
\hline
$No$    &$l$&   $F$    &   $A$  &   $\Phi$ &$S/N$&    $F$   &  $A$  &  $\Phi$  &$S/N$ &   $F$    &  $A$  &  $\Phi$  &$S/N$&   $F$    &  $A$  &  $\Phi$  &$S/N$\\
        &   &  (c/d)   &(mmag)  &($\degr$) &     &   (c/d)  &(mmag)  & ($\degr$)&     &  (c/d)   &(mmag) & ($\degr$)&     &  (c/d]   &(mmag) & ($\degr$)&     \\
\hline
$f_1$	&1  &23.051 (1)&7.3 (3)	&196 (2)   &13.8 &23.054 (1)&5.7 (4) &  166 (4)&8.4   &	  -	     &	   - &	   -    &-	  &23.055 (1)&3.2 (4)&156 (6)	&6.8  \\
$f_2$&1 or 3&15.169 (1)&4.6 (3)	&305 (3)   &6.1	 &15.159 (1)&3.6 (4) &	320 (6)&6.7   &	  -	     &	   - &	   -    &-	  &15.169 (1)&1.7 (4)&312 (12)  &2.4  \\
$f_3$	&-  &17.426 (1)&3.1 (3)	&370 (5)   &5.1	 &17.435 (1)&3.5 (4) &	356 (6)&5.3   &	  -	     &	   - &	   -    & -	  &17.440 (1)&0.9 (4)&334 (24)  &2.0  \\
\hline
        &   &     \multicolumn{4}{c}{CZ Aqr}     &     \multicolumn{4}{c}{UW Cyg}     &    \multicolumn{4}{c}{HZ Dra}     &    \multicolumn{4}{c}{AU Lac}     \\
\hline
        &   &  \multicolumn{4}{c}{$B$-filter}    &   \multicolumn{4}{c}{$B$-filter}   &   \multicolumn{4}{c}{$B$-filter}  &   \multicolumn{4}{c}{$B$-filter}  \\
\hline
$No$    &$l$&   $F$    &   $A$  &   $\Phi$ &$S/N$&    $F$   &  $A$  &  $\Phi$  &$S/N$ &   $F$    & $A$   &  $\Phi$  &$S/N$&   $F$    & $A$   &  $\Phi$  &$S/N$\\
        &   &  (c/d)   &(mmag)  &($\degr$) &     &   (c/d)  &(mmag) & ($\degr$)&     &  (c/d)   &(mmag)  & ($\degr$) &    &  (c/d]   &(mmag) & ($\degr$)&     \\
\hline
$f_1$	&   &35.508 (2)&3.7 (5)	&329 (7)   &8.5	 &27.841 (2)&1.9 (2)&132 (6)   &4.4   &51.068 (2)&	4.0 (4)	&170 (6)& 6.5 &58.217 (1)& 5.0 (3)& 177 (3)	&11.9 \\
$f_2$	&   &	-      &  -	    &	  -	   &-	 &32.087 (2)&1.7 (2)&111 (6)   &4.1   &     -	 &	  -	    &	  -	&  -  &51.298 (1)& 4.1 (3)&	356 (4)	&4.0  \\
\hline
\end{tabular}}
\end{table}
\end{landscape}

\section{Evolution and pulsation-period correlations}

Seventy four cases of close binaries containing a $\delta$~Sct component are listed in Table~8. In particular, the columns contain: the name of the \emph{System}, its orbital period $P_{\rm orb}$, its dominant pulsation period $P_{\rm puls}$ and its semi-amplitude \emph{A}, the mass $M$ and the radius $R$ of the pulsating component, the geometrical configuration of the system \emph{Type} and the corresponding literature reference \emph{Ref}.

Three new correlation diagrams (Fig.~6) between pulsation and orbital periods are illustrated. Since the range of orbital periods is larger than the one presented by \citet{SO06a}, we chose to correlate logarithmic rather than decimal values. For the following derived relations all except four systems were used. In particular, CPD-31$\degr$6830, CPD-41$\degr$5106 and CPD-60$\degr$871 \citep{PI07} are ambiguous concerning both their geometrical types and pulsational characteristics, while V1241~Tau (referred also as WX~Eri) was listed as an oEA system by \citet{SO06b}, but \citet{AR04} found no evidence of pulsations. So far there is no clear theoretical relation between orbital and pulsation periods. Therefore, systems were distinguished according to their geometrical types and an empirical relation for each subset was extracted:\\
For \textsl{Semidetached} binaries (oEA systems):
\begin{equation}
\log P_{\rm puls}= -1.56(4)+0.62(8)~\log P_{\rm orb}
\end{equation}
for \textsl{Detached} binaries:
\begin{equation}
\log P_{\rm puls}= -1.4(1)+0.5(2)~\log P_{\rm orb}
\end{equation}
for \textsl{all} known close binaries (Detached, Semidetached, unclassified) including a $\delta$~Sct component:
\begin{equation}
\log P_{\rm puls}= -1.53(3)+0.58(7)~\log P_{\rm orb}
\end{equation}

\begin{figure}
\centering
\begin{tabular}{c}
\includegraphics[width=8cm]{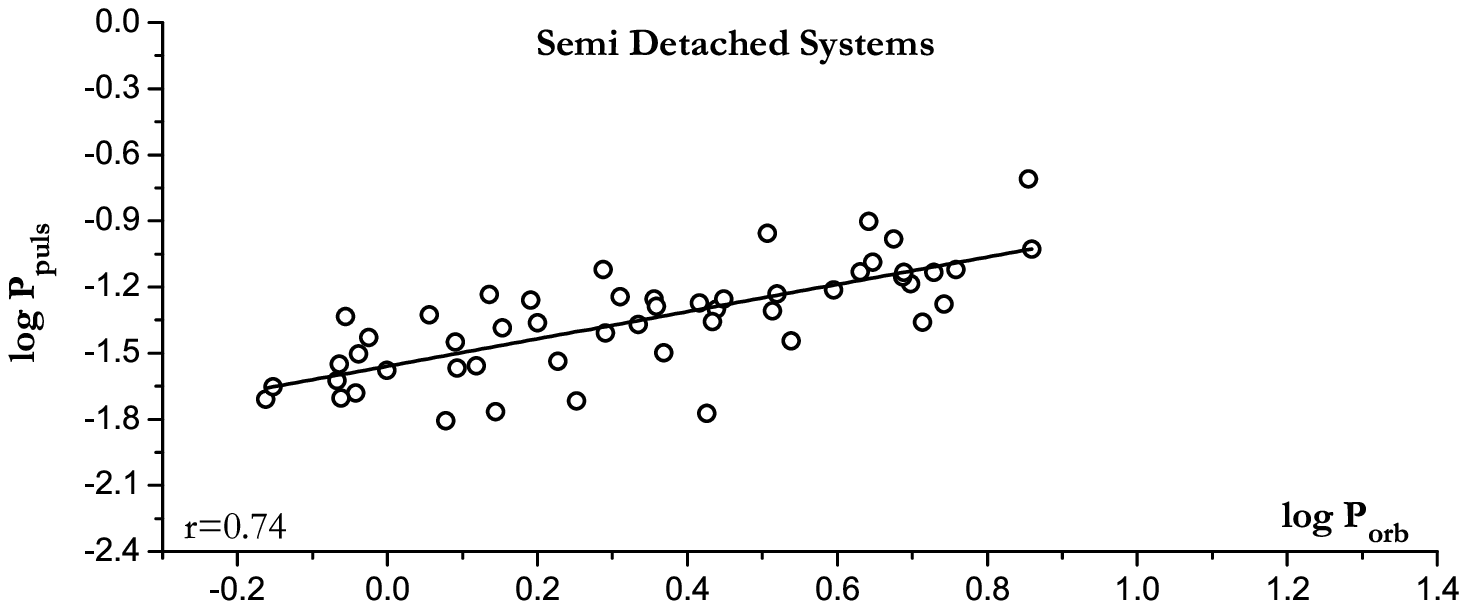}\\
\includegraphics[width=8cm]{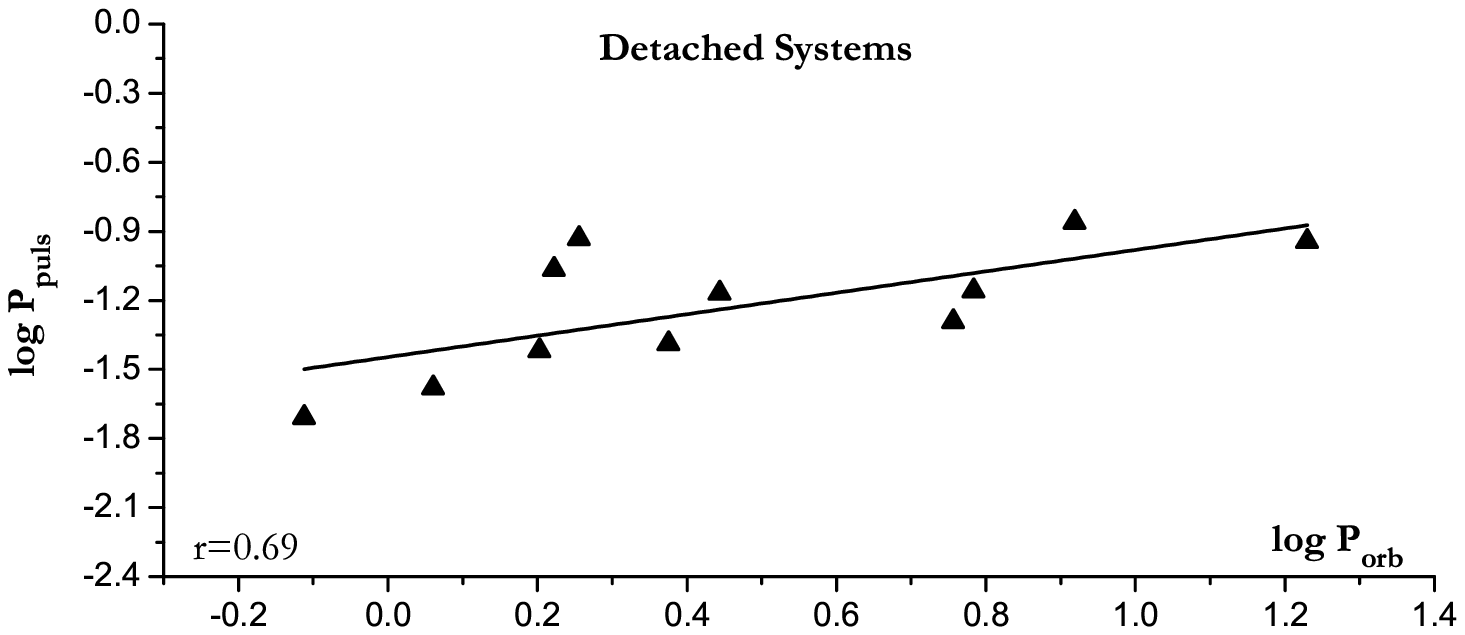}\\
\includegraphics[width=8cm]{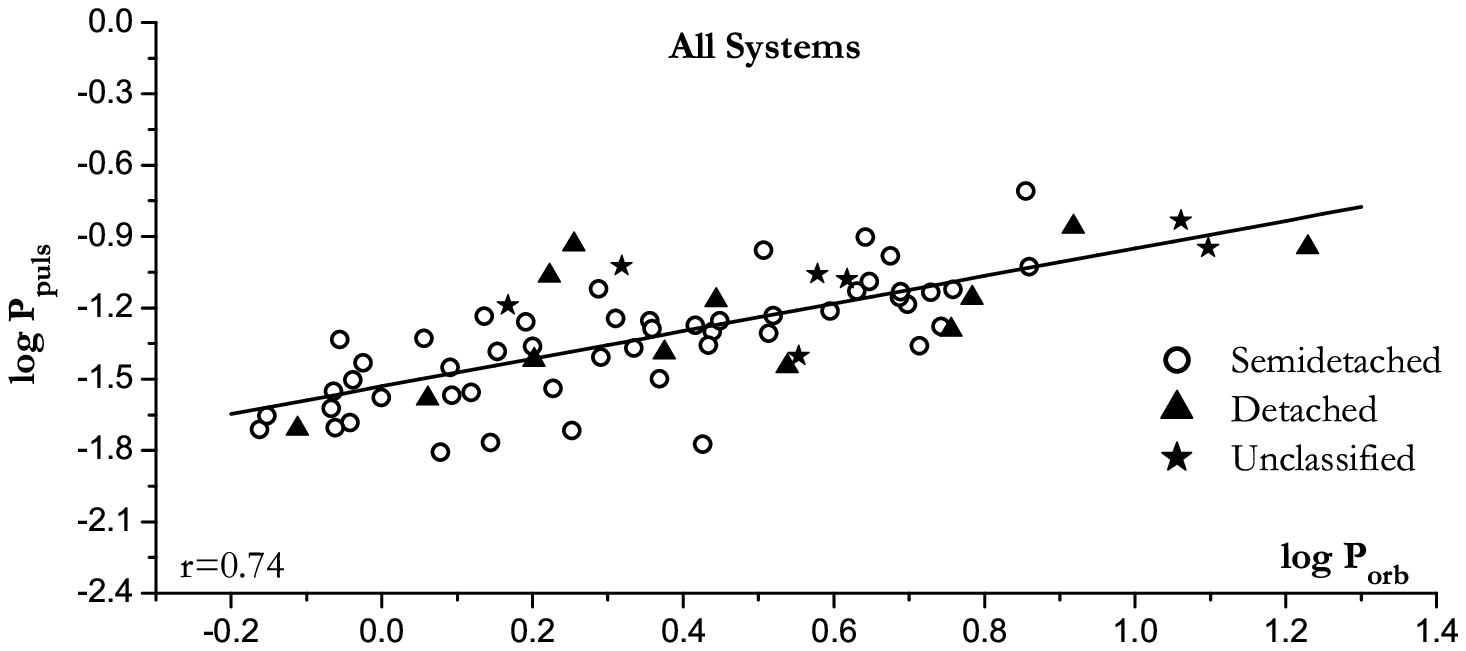}
\end{tabular}
\caption{The correlation between pulsation and orbital periods for $\delta$~Sct stars in semidetached, detached and all close binaries. Black solid lines represent the best linear fittings and the correlation coefficient $r$ is also indicated.}
\label{fig6}
\end{figure}

Although the sample of the present paper is almost quadruple than that of \citet{SO06a}, it is still not large (70 cases included). From our findings, there is no unique physical law correlating pulsation and orbital periods for all kind of binaries containing a $\delta$~Sct member. However, the results of the present study strongly suggest that there are some connections between these quantities and further theoretical investigation is needed.

Using the mass and radius of a pulsating component, we can calculate the surface gravity $g$, which can be related to the stage of stellar evolution. It was noticed that the more evolved the star the more slow the pulsations, so a connection between its pulsation period and evolutionary status is plausible. This may be similar to that given for radially pulsating stars \citep{FE95}:
\begin{equation}
\log g= -1.14(4)~\log P_{\rm puls} + 2.62(4)
\end{equation}
For the correlation between $\log g$ and $P_{\rm puls}$ our sample consisted of systems with known absolute parameters (Table~8), for which most are semi-detached ones (41 of 46).
A 3D plot presenting the correlation between dominant pulsation frequency and evolutionary status of binary members-$\delta$~Sct stars is given in Fig.~7, where it is shown that the faster pulsating stars are inside ZAMS-TAMS limits, while the slower ones lie beyond TAMS. Their positions in the $M-R$ diagram (i.e. the $x-y$ level of Fig.~7) are shown in Fig.~8. The dominant pulsation periods of these stars are plotted against their $\log g$ in Fig.~9. The following linear function was found to fit the data adequately well:
\begin{equation}
\log g= -0.3(1)~\log P_{\rm puls} + 3.7(2)
\end{equation}

Although this relation is very different to that of single pulsating stars, as shown in Eq.~4-5, it can be noticed that the scatter of the points is large. Especially the data set between $-1.4<\log P_{\rm puls}<-1.0$ shows the largest deviation from the theoretical trend, while for $\log P_{\rm puls}>-1.0$ there are only a few points. The possibility of two branches existence in the data set was examined, but since the sample consists of stars of the same type (i.e. oEA stars of similar evolutionary stage) we could not find any physical interpretation in presenting different correlations between $P_{\rm puls}$ and $g$ according to the $P_{\rm puls}$ values. However, a larger sample of oEA stars in the future would clarify this hypothesis.

\section[]{Discussion and conclusions}

68 eclipsing systems candidates for pulsations were observed, in order to check them for possible short-periodic oscillations. Eight new systems (QY~Aql, CZ~Aqr, UW~Cyg, HL~Dra, HZ~Dra, AU~Lac, CL~Lyn and IO~UMa) were discovered and their dominant pulsation characteristics listed. These findings increase the current sample of binaries with $\delta$~Sct components by $\sim$11\%.

\begin{landscape}
\begin{table}
\caption{List of close binaries containing a $\delta$~Sct component.}
\scalebox{0.93}{
\begin{tabular}{l cc cc cc c l cc cc cc c}
\hline
System   &$P_{\rm orb}$&$P_{\rm puls}$&    $A$   &   $M$    & $R$&Type&  Ref.&  System  &$P_{\rm orb}$&$P_{\rm puls}$&  $A$     &  $M$   &  $R$ &Type&Ref.    \\
         &    (d)      &       (d)    & (mmag)   &($M_{\sun}$)&($R_{\sun}$)&&&          &         (d) &     (d)      & (mmag)   &($M_{\sun}$)&($R_{\sun}$)&&  \\
\hline
Aql QY	 &  7.22954    &    0.09385   &11.9 ($B$)&     --   &  --  &SD&pp    & HD 99612 &	2.77876	    &	0.06796	 &	  --    &	 --	 &  --  &D  & 31      \\
Aqr CZ   &	0.86275	   &	0.02849	  &	4.0 ($B$)&	 2.0    &	1.9&SD&pp    &HD 172189 &	5.70165	    &	0.05100	 &	  --    &	 --	 &  --  &D  &35, 7    \\
Aqr DY   &	2.15970	   &	0.04275	  &13.0 ($V$)&	--      &	-- &SD&37, 43& HD 207651&	1.47080	    &	0.06479	 &21.4 ($B$)&	 --  &  --  &U  &19       \\
Aur KW   &	3.78900	   &	0.08750	  &80.0 ($V$)&	  2.3	&	4.0&U &20, 2 & HD 220687&	1.59425	    &	0.03821	 &12.8 ($V$)&	 --  &  --  &D  &31	      \\
Boo EW   &	0.90630	   &	0.02083	  &20.0 ($V$)&	 --     &	-- &SD&36	 &   Her BO &	4.27283	    &	0.07446	 &68.0 ($B$)&	2.4	 &	3.6	&SD&39, 3, pp \\
Boo YY   &	3.93307	   &	0.06128	  &58.4 ($B$)&	2.0	    &  1.9 &SD&17, 41&   Her CT &	1.78640	    &	0.01889	 & 3.2 ($B$)&	2.3	 &	2.1	&SD&34, 41, 25\\
Cam Y    &	3.30570	   &	0.05860	  &11.6 ($V$)&	1.7	   &2.9&SD&36, 22, 34&  Her EF  &	4.72920	    &	0.10420	 &60.0 ($B$)&	1.5	 & 1.6  &SD	&	34, 41\\
CMa R    &	1.13590	   &	0.04710	  &8.8 ($B$) &	1.1	    &  1.5 &SD&34, 35&   Her TU	&	2.26690	    &	0.05560	 &8.0 ($V$)	&	1.4	 & 1.6  &SD	&34, 40	  \\
Cap TY   &	1.42346	   &	0.04132	  &21.0 ($B$)&	2.0	    &2.5   &SD&27, pp&  Her V944&	2.08309	    &	0.09467	 &31.0 ($V$)&	-- 	 &	 -- &U	&9	      \\
Cas AB   &	1.36690	   &	0.05830	  &39.2 ($V$)&	2.3	    &2.0   &SD&34, 35& HIP 7666 &	2.37232	    &	0.04090	 &20.0 ($V$)&	-- 	 &	 -- &D	&	34    \\
Cas IV   &	0.99852	   &	0.02650	  &10.0 ($B$)&	2.6	    &2.0   &SD& 35	 &  Hya AI 	&	8.28970	    &	0.13800	 &20.0 ($V$)&	2.0	 &	 1.8&D	&	34    \\
Cas RZ   &	1.19530	   &	0.01560	  &	13.0 (Y) &	2.3	    &1.6   &SD&34, 35&  Hya RX  &	2.28170	    &	0.05160	 &14.0 ($B$)&	1.7	 &	 1.7&SD	&34, 22	  \\
Cep XX   &	2.33732	   &	0.03174	  &	3.8 ($B$)&	2.0	    &2.1   &SD&29, 35&KIC 10661783&	1.23136	    &	0.03554	 &   4.0	&	 --  &	 -- &SD	& 33	  \\
Cet WY   &	1.93969	   &	0.07575	  &	8.6 ($B$)&	1.7	    &2.2   &SD&27, pp&   Lac AU &	1.39243	    &	0.01977	 &7.2 ($B$) &	2.0	 &	 1.8&SD	&  pp     \\
Cha RS   &	1.66987	   &	0.08600	  &	16.8	 &	1.9	    &2.2   &D &  35	 &   Leo DG &	4.14675	    &	0.08337	 &6.2 ($B$) &   --   &	 -- &U	&  24	  \\
CPD-$31\degr6830^a$ &0.88343&0.18304  &54.1 ($V$)&	 --     &   -- &D &	 31	 &   Leo WY	&	4.98578	    &	0.06550	 &11.0 ($V$)&	2.3	 &	 3.3&SD	&16, 3    \\
CPD-$41\degr5106^a$ &2.13700&0.12125  &20.2 ($V$)&	 --     &   -- &D &	 31	 &   Leo Y  &	1.68610	    &	0.02900	 &4.1 ($V$) &	1.6	 &	 1.7&SD	&42, 40	  \\
CPD-$60\degr871^a$  &1.22096&0.21423  &17.1 ($V$)&	 --	    &	-- &D &	 31	 &   Lep RR &	0.91543	    &	0.03138	 &5.0 ($V$) &	2.2	 &	 2.0&SD	&16, 41	  \\
Cyg UW   &	3.45080	   & 0.03405	  &	4.2 ($B$)&	1.9	    &	2.3&SD&	 pp  &   Lyn CL &	1.58604	    &	0.04338	 &7.3 ($B$) &	2.0	 &	 2.5&SD	&	pp    \\
Cyg V346 &	2.74330	   & 0.05020	  &30.0 ($B$)&	2.3	    &3.8&SD&23, 34, 3&  Lyn CQ  &	12.50736    &	0.11277	 &40.0 ($V$)&	-- 	 &	 -- &U	&	4     \\
Cyg V469 &	1.31250	   &	0.02780	  &20.0 ($V$)&	3.3	    &	2.7&SD&34, 41&   Mic VY &	4.43637	    &	0.08174	 &19.4 ($V$)&	2.4	 &	 2.2&SD	&	31, 3 \\
Dra GK   &	16.96000   &	0.11376	  &40.0 ($V$)&	 --     &	-- &D &	  8	 & Oph V577 &	6.07910	    &	0.06950	 &28.9 ($V$)&	1.7	 &	 1.8&D	&	35, 41\\
Dra HL   &	0.94428	   &	0.03848	  &34.4 ($B$)&	2.5	    &	2.5&SD&	pp   & Oph V2365&	4.86560	    &	0.07000	 &50.0 ($V$)&	2.0	 &	 2.2&SD	&	21    \\
Dra HN   &	1.80075	   &	0.11686	  &10.9 ($B$)&	 --     &	-- &D &	 5	 &	Ori FL  &	1.55098	    &	0.05501	 &44.0 ($V$)&	2.9	 &	 2.1&SD	&	43, 3 \\
Dra HZ   &	0.77294	   &	0.01895	  &51.1 ($B$)&	3.0	    &	2.3&D &	pp   & Pav MX  	&	5.73084	    &	0.07560	 &76.9 ($V$)&	-- 	 &	 -- &SD	&	30	  \\
Dra SX   &	5.16957	   &	0.04375	  &4.0 ($V$) &	1.7	    &	3.1&SD&15, 40& Peg BG  	&	1.95243	    &	0.04002	 &15.0 ($V$)&	2.2	 &2.0&SD&37, 16, 38, 28\\
Dra TW   &	2.80690	   &	0.05560	  &10.0 ($B$)&	1.6	    &	2.4&SD&34, 35& Per AB  	&	7.16030	    &	0.19580	 &20.0 ($B$)&	1.9	 &	 2.0&SD	&34, 40	  \\
Dra TZ   &	0.86600	   &	0.01973	  &	5.6 ($B$)&	2.1	    &	2.0&SD&	34, 3& Per IU  	&	0.85700	    &	0.02380	 &20.0 ($B$)&	2.4	 &	 1.9&SD	&  34, 3  \\
Eri AS   &	2.66410	   &	0.01690	  &	6.8 ($V$)&	1.9	    &1.6   &SD&	  34 & Pyx XX  	&	1.15000	    &	0.02624	 &10.1 ($B$)&   -- 	 &	 -- &D	&	1	  \\
Eri TZ   &	2.60620	   &	0.05340	  &83.0 ($B$)&	2.0	    &	1.7&SD&34, 26& Ser AO  	&	0.87930	    &	0.04650	 &20.0 ($B$)&	2.6	 &	 1.8&SD	&	34    \\
Gru RS   &	11.50000   &	0.14700	  &600.0 ($B$)&	 --     &	-- &U &	10	 & Tau AC  	&	2.04340	    &	0.05703	 &6.0 ($V$) &	 1.5 &	 2.3&SD	&	16, 3 \\
GSC 3889-0202&2.71066  &	0.04410	  &50.0 ($V$)&	 --     &	-- &SD&	12	 &Tau V1241$^{a,b}$&0.82327	&	0.16450	 &   30	    &	1.7	 &	 1.9&SD	&	35    \\
GSC 4293-0432&4.38440  &	0.12500	  &40.0 ($B$)&	 --     &   -- &SD&	14	 & Tel IZ  	&	4.88022	    &	0.07376	 &45.9 ($V$)&   -- 	 &	 -- &SD	&	31    \\
GSC 4550-1408&1.23837  &	0.02703	  &20.0 ($B$)&	 --     &   -- &SD&	11	 &UNSW-V-500&	5.35048	    &	0.07340	 &350.0 ($V$)&	 1.5 &	 2.4&SD	&	6	  \\
GSC 4558-0883&3.25855  &	0.04930	  &15.0 ($R$)&	 --     &	-- &SD&	13	 &	UMa IO  &	5.52039	    &	0.05275	 &6.7 ($B$) &	-- 	 &	 -- &SD	&	pp    \\
HD 61199 &	3.57436	   &	0.03959	  &1.5 ($V$) &	 --     &   -- &U &	18	 &	UMa VV  &	0.68740	    &	0.01950	 &15.0 ($B$)&	2.3	 &	 1.7&SD	& 34, 35  \\
HD 62571 &	3.20865	   &	0.11048	  &41.7 ($V$)&	 --     &   -- &SD&	31	 &	Vel BF 	&	0.70400	    &	0.02225	 &26.0 ($B$)&	2.0	 &	 1.8&SD	&	30	  \\
\hline
\end{tabular}}
D=Detached, SD=Semidetached, U=Unclassified, $^a$Not well defined, $^b$WX~Eri~~~~~~~~~~~~~~~~~~~~~~~~~~~~~~~~~~~~~~~~~~~~~~~~~~~~~~~~~~~~~~~~~~~~~~~~~~~~~~~~~~~~~~~~~~~~~~~~~~~~~~~~~~~~~~~~~~~~~~~~~~~~~~~~~~~~~~~~~ \\
Ref.: pp=present paper; (1) \citet{AE02}; (2) \citet{AL99}; (3) \citet{BU04}; (4) \citet{CA02}; (5) \citet{CH04}; (6) \citet{CH07}; (7) \citet{CO07}; (8) \citet{DA02}; (9) \citet{DM05}; (10) \citet{DE09}; (11) \citet{DI08a}; (12) \citet{DI08b}; (13) \citet{DI09a}; (14) \citet{DI09b}; (15) \citet{DI10}; (16) \citet{DV09}; (17) \citet{HA10}; (18) \citet{HA08}; (19) \citet{HE04}; (20) \citet{HU71}; (21) \citet{IB08}; (22) \citet{KI03}; (23) \citet{KI05}; (24) \citet{LA05}; (25) \citet{LA11}; (26) \citet{LI08}; (27) \citet{LN09}; (28) \citet{LN11}; (29) \citet{LE07}; (30) \citet{MA09}; (31) \citet{PI07}; (32) \citet{RO10}; (33) \citet{SOU11}; (34) \citet{SO06a}; (35) \citet{SO06b}; (36) \citet{SO08}; (37) \citet{SO09}; (38) \citet{SO11}; (39) \citet{SB07}; (40) \citet{SS04}; (41) \citet{SK90}; (42) \citet{TU08}; (43) \citet{ZA11}~~~~~~~~~~~~~~~~~~~~~~~~~~~~~~~~~~~~~~~~~~~~~~~~~~~~~~~~~~~~~~~~~~
\end{table}
\end{landscape}

\begin{figure}
\begin{tabular}{c}
\includegraphics[width=7.2cm]{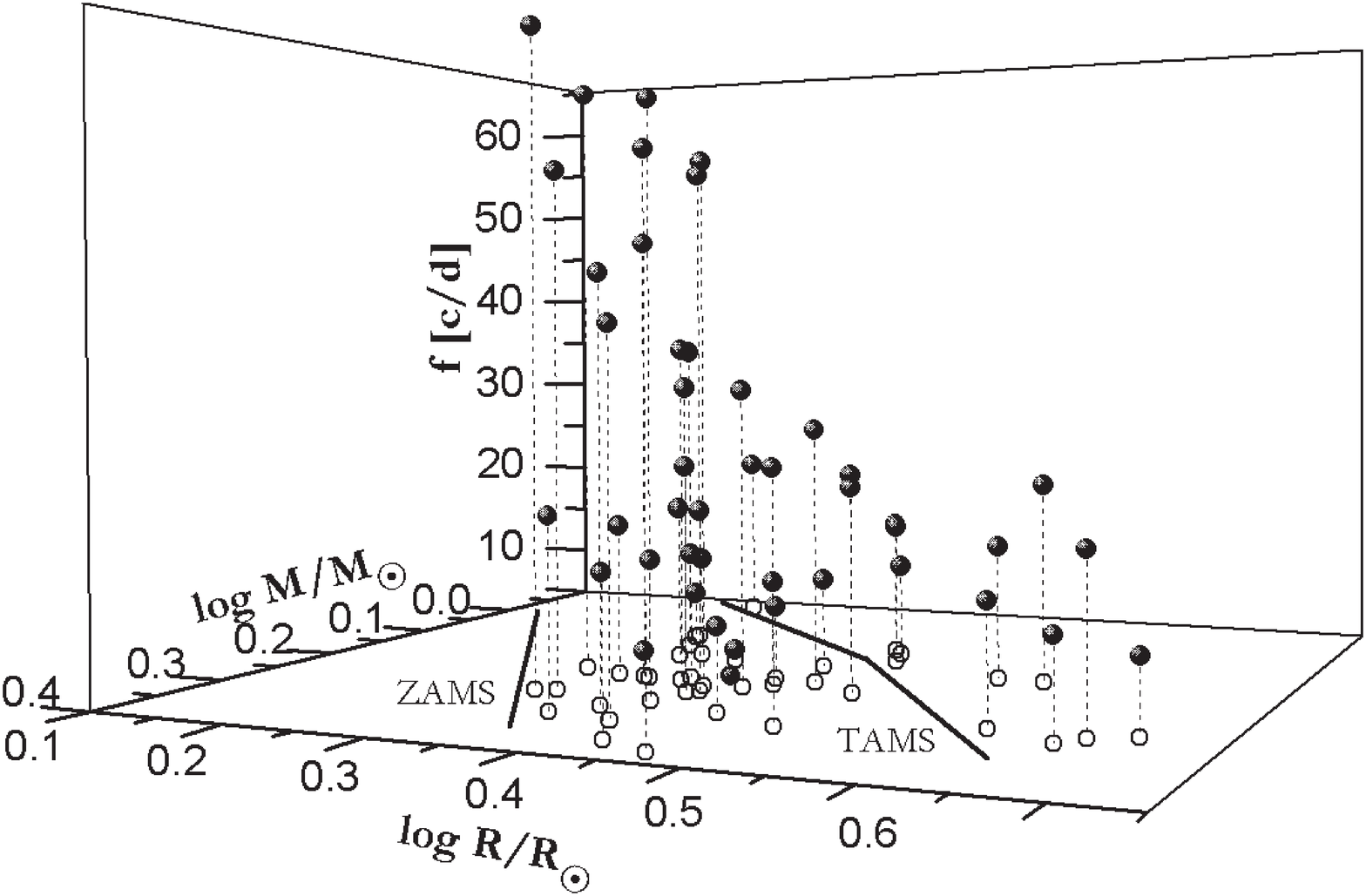}\\
\end{tabular}
\caption{The 3D correlation plot of the evolutionary stage ($x-y$ level) and the dominant frequency ($z$-axis) for $\delta$~Sct components in close binaries. The filled symbols represent the data points, while their projection in the $M-R$ diagram is indicated by dash lines guiding into their respective open circles.}
\label{fig7}

\begin{tabular}{c}
\includegraphics[width=8cm]{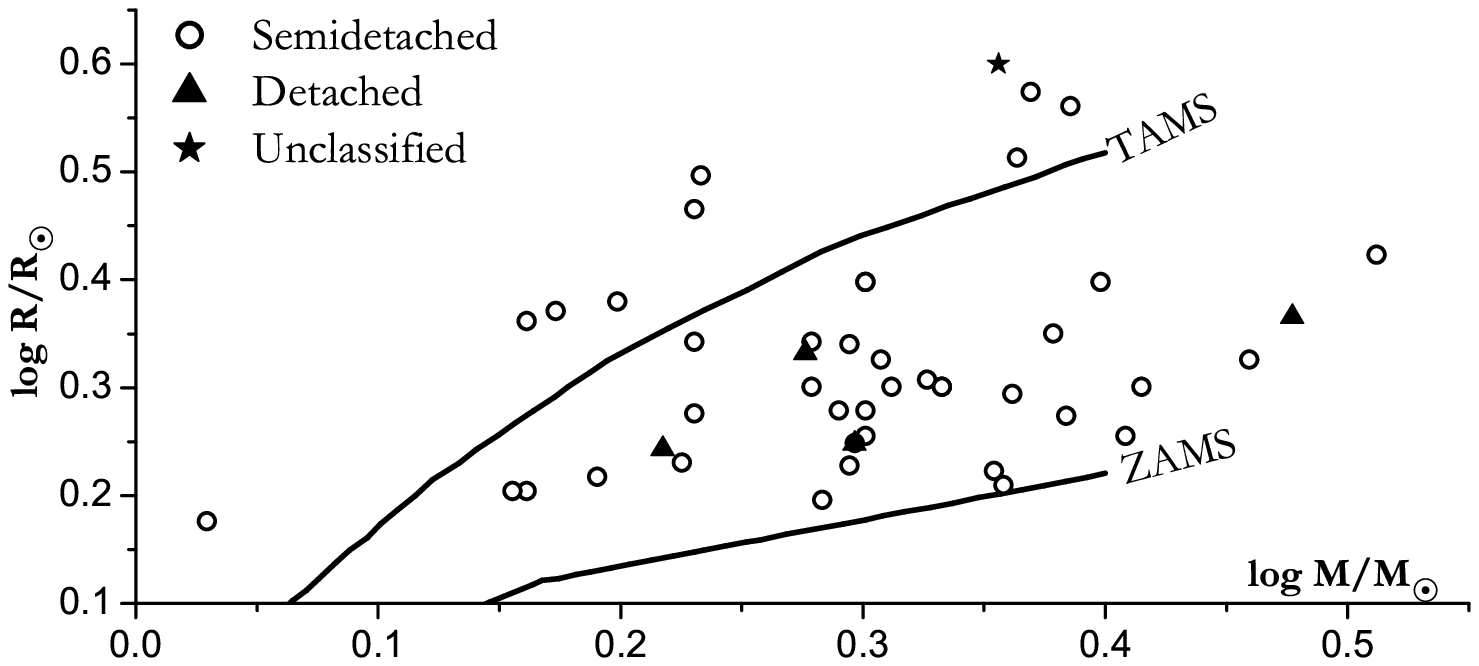}\\
\end{tabular}
\caption{The location of binary members $\delta$ Sct stars in the $M-R$ diagram. The ZAMS and TAMS limits are drawn with black solid lines.}
\label{fig8}

\begin{tabular}{c}
\includegraphics[width=8cm]{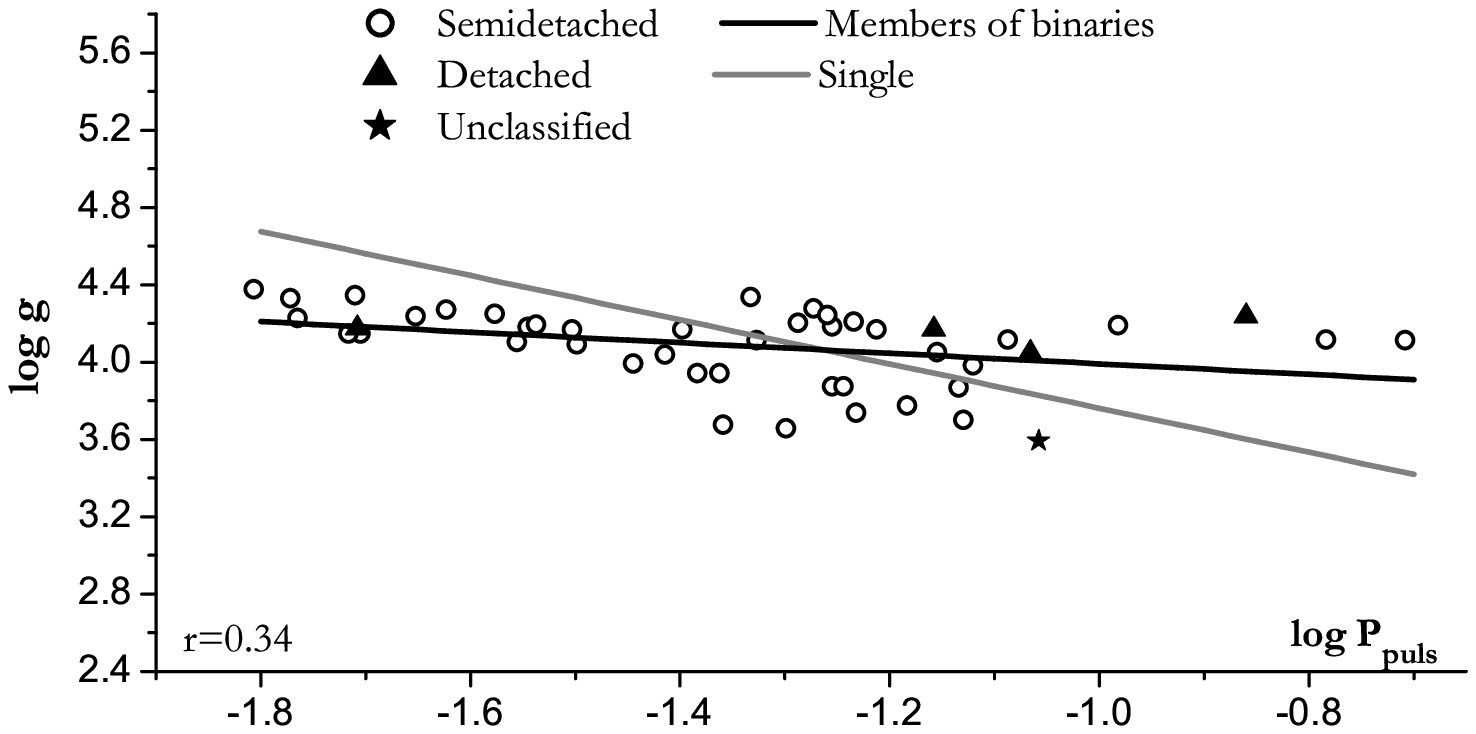}\\
\end{tabular}
\caption{The gravity acceleration versus the dominant pulsation frequency for the $\delta$~Sct stars-members of close binaries (symbols). The black solid line shows the best linear fit, while the grey one corresponds to the respective relation for single $\delta$~Sct stars \citep{FE95}.}
\label{fig9}
\end{figure}

The systems V821~Cas, EI~Cep, CM~Lac, DM~Peg, UZ~Sge and RR~Vul are still candidates, since they presented some `suspicious' frequencies, but within the error limits of the measurements. Further observations with larger telescopes are needed to check these candidates. The systems: SZ~Ari, V417~Aur, V364~Cas, MY~Cyg, V477~Cyg, SZ~Her, UX~Her, SX~Lyn, EP~Mon, EY~Ori, FT~Ori and V536~Ori were re-observed in this way, but the results were negative \citep[see also][]{DV09}.

LCs of eight systems including an oscillating component were analyzed in order to derive their physical, geometrical and pulsational characteristics. A semi-detached configuration turned out for all systems, except for HZ~Dra that was found to be detached. Their primaries were identified as MS $\delta$~Scuti-like pulsators, therefore, according to the definition suggested by \citet{MK04}, all systems, except for HZ~Dra, can be considered as oEA stars.

For CZ~Aqr evidence supporting mass transfer was not detected in the O$-$C analysis, neither was any third light indicated in the LC fittings. The mass transfer might be at a starting/ending phase, so that the rate could be too small to be detected. The third body, according to its minimal mass and assuming its MS nature, should contribute less than 1\% to the total light, therefore its non-detection is reasonable.. On the other hand, the quadrupole moment of the secondary component (1.4$\times10^{50}$~g~cm$^2$) can also account for cyclic orbital period changes. One pulsation frequency with relatively small amplitude was found for the primary component.

Mass transfer from the secondary to the primary component of TY~Cap is supported by both LC and O$-$C analyses. The detected third light had a value of $\sim$3\%, while the minimal mass found for the tertiary component (0.64~$M_{\sun}$) suggests $\sim$1\% light contribution to the total light, as can be derived from the Mass-Luminosity relation ($L\sim M^{3.5}$) for the MS stars. Therefore, we conclude that these results are mutually supportive for the triplicity of the system. Alternatively, the 4.4$\times10^{50}$g~cm$^2$ value of $\Delta Q$ could be consistent with a possible Applegate-type mechanism. For the MS component of the system two oscillation frequencies were detected.

For WY~Cet the O$-$C and LC analyses are consistent with mass flow from the less to the more massive component, while the detected third light was found to be $\sim$1\%. The light contribution of a MS star having the derived minimal mass is expected to be $\sim$4\%. This disagreement questions the existence of the third body or its possible MS nature. Plausibly, observed cyclic changes of the EB's period can be explained as magnetic influences of the secondary component ($\Delta Q=3.2\times10^{50}$~g~cm$^2$). The frequency analysis resulted in one oscillating mode for the system's primary component.

The O$-$C analysis of UW~Cyg yielded two cyclic orbital period variations, while the detected additional light was found $\sim$1\%. The values of quadrupole moment's variation of the secondary component were found to support both orbital period's cyclic changes. On the other hand, the hypothetical fourth body was found to have a minimal mass of $\sim$0.6~$M_{\sun}$, which can explain the observed additional light contribution, assuming its MS nature. To sum up, the most possible explanation for the double periodic behaviour of the system's orbital period seems to be the secondary component's magnetic quadrupole moment variation with a period of about 82~yr and the existence of a tertiary companion with a period of $\sim$30~yr. No parabolic term in the O$-$C solution was found, showing that the mass transfer rate is too weak to be detected in the current data set. For the primary component two low amplitude frequencies were traced.

For HZ~Dra we found one, for HL~Dra and CL~Lyn three and for AU~Lac two pulsationfrequencies.

Taking into account only the reliable cases of binaries containing a $\delta$~Sct star, three new empirical relations connecting pulsation and orbital periods are proposed. These relations are certainly not the final ones due to the rather small sample. In coming years it is expected that the sample will be enriched significantly and more detailed relationships between pulsation and orbital periods will be tested. The pulsating members of oEA systems, as distinct from the ones in detached, are mass accreting and the mass flow on their surface should affect the pulsations \citep{MK07}. Our results showed that the $\delta$~Sct stars in near contact systems present slightly slower pulsations than the ones in detached binaries for a given orbital period.

The majority of the binary members-$\delta$~Sct stars lie inside the MS band and closer to the ZAMS limit, indicating that their instability starts at early stages of their evolution in comparison with the singles. Another significant conclusion is the relation connecting the evolutionary stage of a $\delta$~Sct star member of a close binary with its dominant pulsation frequency. A closer examination of this, reveals that the older the star the slower the pulsations. Although this is already known for single pulsators, a new correlation between evolutionary stage and dominant pulsation frequency for binary-members $\delta$~Sct stars is introduced in the present study, which was found to differ significantly from that for the respective singles. 80\% of $\delta$~Sct in binaries pulsate with periods shorter than 2~hr \citep{SO06b}, while the majority of single $\delta$~Sct stars (68\%) have periods in the range of 1.2-3.6~hr \citep{ROD00}. Clearly, pulsating members of binaries follow a different evolutionary track than classical ones as it was firstly pointed out by \citet{MK03}. The current sample includes mostly mass gaining $\delta$~Sct stars and it seems that the mass accretion plays an essential role in the star's pulsation frequency decrease rate, in a way that the star pulsates faster than if it was single for a given evolutionary stage.

Future discoveries of binaries containing $\delta$~Sct components will increase the current sample to better test possible dependence of the pulsational behaviour on the mass exchange. Moreover, a physical interpretation of the empirical relations correlating the pulsation and orbital periods, and the pulsation period with the evolutionary stage of the $\delta$~Sct components in close binaries is certainly needed. Future theoretical modelling for mass gaining $\delta$~Sct stars should take into account quantities such as mass, radius, mass transfer rate, radius expansion rate and dominant pulsation period. Hence, the combination of theoretical models and current observational results is expected to answer a lot of open questions, enriching our knowledge about stellar evolution near the end of the Main Sequence phase.

\section*{Acknowledgments}

This work has been financially supported by the Special Account for Research Grants No 70/4/11112 of the National \& Kapodistrian University of Athens, Hellas for A.L., and the Czech Science Foundation grant no. P209/10/0715 for P.Z. Skinakas Observatory is a collaborative project of the University of Crete, and the Foundation for Research and Technology-Hellas. We thank E. Budding for carefully reading the manuscript and his essential corrections, and the anonymous referee for the valuable comments and suggestions which improved substantially the paper. In the present work, the minima database: (http://var.astro.cz/ocgate/) has been used.

\bsp
\label{lastpage}

\end{document}